\documentclass[11pt]{article}
\usepackage{graphicx}
\usepackage{amssymb}
\usepackage{amsmath}
\usepackage{geometry}
\usepackage{xcolor}
\usepackage{verbatim}
\usepackage{enumitem}
\usepackage{ulem}
\usepackage{float}
\usepackage{bm}
\usepackage{natbib}
\usepackage{setspace}
\usepackage{amsfonts}

\usepackage{appendix}
\numberwithin{equation}{section} 

 \usepackage{mdframed}

\setcitestyle{authoryear,open={(},close={)}}

\begin{document}
\sloppy

\begin{center}
{\Large \textbf{Discrete vs. continuous dynamics in biology: }}

\medskip

{\Large \textbf{Structural parallels and differences}}

\bigskip

Shuyun Jiao$^{1}$ and David Waxman$^{2}$

\bigskip

$^{1}$Shanxi Key Laboratory of Cryptography and Data Security.

Shanxi Normal University, 339 Taiyu Road, Taiyuan, Shanxi, 030031, PRC

\medskip

$^{2}$Centre for Computational Systems Biology, ISTBI,

Fudan University, 220 Handan Road, Shanghai 200433, PRC
\end{center}

\medskip

\noindent$^{1}$Email: jiaosy@sxnu.edu.cn \hfill ORCID ID: 0000-0002-4738-4668

\noindent$^{2}$Email: davidwaxman@fudan.edu.cn \hfill ORCID ID: 0000-0001-9093-2108

\noindent$^{2}$Corresponding author.

\begin{center}
\textbf{{\Large {Abstract}}}
\end{center}

Biological systems exhibiting discrete-time dynamics are governed by
difference equations. Examples arise in population genetics, ecology,
evolution, when generations are non-overlapping, and also in epidemic
modelling, when numbers of infected individuals are recorded at fixed intervals.

In this work we consider time-homogeneous and time-inhomogeneous problems. 
Exactly soluble difference equations are shown to have exact continuous-time 
descriptions which, at the discrete times, precisely reproduce the discrete-time 
solution. Such descriptions work for arbitrarily large changes from one discrete 
time to the next.

We extend this approach to problems without exact solutions by employing
approximate mappings between difference and differential equations. We produce
approximate continuous-time models that to high accuracy capture extremely 
rapid growth and decay.

The mapping from discrete to continuous time exposes structural parallels and
differences. For example, some coefficients are constant within the differential equation but 
are paralleled by appreciably non-constant functions within the difference equation.

Difference equations can sometimes have solutions containing rapid oscillatory behaviour, namely a sign change each time step. We show some exactly soluble cases have a continuous-time description, at the expense of being complex-valued.

Bridging discrete and continuous representations of a system enhances insights, analysis and approximation of the dynamics. 

\bigskip


\newpage
\section{Introduction}

Discrete-time dynamics arise naturally in biological systems. Classical
examples include: annual plant populations with non-overlapping generations
\citep{Maynardsmith}; disease surveillance data collected at fixed intervals
\citep{Yakubu,earn2000}; seasonal diseases such as measles
\citep{earn2000}, which can exhibit near-periodic variation;
pharmacokinetic/pharmacodynamic modelling for discrete dosing
\citep{mager2003diversity}. Each of these applications illustrates biological
phenomena that involve discrete or near discrete events, possibly within the
context of continuous-time systems.

Discrete-time dynamics is usually described by a \textit{difference equation}.
A naive continuous-time approximation follows by first approximating the
difference equation by a \textit{differential equation}, and then finding the
continuous-time solution. Such an approximation is plausibly used when changes
between discrete time steps are small. However, some biological systems can
exhibit appreciable changes between time steps, e.g., viral load doubling, the
rapid spread of a large $R$ infection, and geometric growth with an
appreciable growth factor. Generally, continuous-time approximations
\textit{fail} for systems with large per-step changes, limiting their utility
in real-world applications like epidemiology or genetics.

In this work we first establish an exact correspondence between the solutions
and equations of some \textit{exactly soluble} discrete-time systems and the
mathematically distinct equations/solutions of some continuous-time systems.
An outcome is that a difference equation describing a discrete-time system
becomes replaced by a differential equation, whose solution \textit{precisely
agrees}, at the discrete times, with the discrete-time solution. This
agreement holds irrespective of the magnitude of the per-step changes.

We then go beyond exactly soluble discrete-time problems. We establish an
approximate correspondence between discrete-time and continuous-time systems. This
amounts to an approximation scheme for discrete-time problems that are not
exactly soluble. The approximation is expressed in terms of the solution of a
differential equation, but now the solutions only approximately agree.

The above approach, which derives either an exact or an approximate
differential equation to describe a discrete-time system, is, as we shall
show, often able to handle extreme parameter values that lead to large
inter-step changes. This bypasses the need for restrictive small-parameter
assumptions for validity of the differential equation. A further feature of
the above approach is that it works for both time-homogeneous and time-inhomogeneous problems. 
This then becomes relevant to some real-world discrete-time cases with 
time-dependent parameters (e.g., seasonal disease transmission).

For completeness, we also consider a class of discrete-time systems that
contain the most rapid oscillatory behaviour possible in such systems. This is
where the dynamical quantity changes sign at every time step. While it would
seem impossible to describe such systems in continuous time, we explicitly
show, in exactly soluble problems, that a continuous-time representation,
involving a differential equation, is possible. However the solutions are
generally complex-valued - with the exception of the discrete times - where
the solutions are real and there is complete agreement with the discrete-time solution.

This work non-trivially relates some discrete-time and continuous-time problems,
and shows how to exploit this relationship in `close' cases. It also shows how
to deal with the issue of rapidly changing solutions in discrete time.

The relationship between discrete-time and continuous-time models arises in many
biological contexts. In population dynamics, both discrete-time and
continuous-time descriptions are routinely used
\citep{caswell2001matrix,kot2001elements} and in epidemiology, both discrete-time and
continuous-time formulations are used
\citep{earn2000,Yakubu,scutt2022theoretically}. 

More generally,
relationships between the steady states of continuous-time and discrete-time
models arising from biological systems have been studied
\citep{VelizCuba2012}. Several mathematical approaches address the relation
between such models. A continuous-time model may be discretised using
standard numerical schemes or through nonstandard finite-difference
constructions designed to preserve particular properties of the continuous
model \citep{Mickens1994}. Time-scale calculus
\citep{BohnerPeterson2001} provides a unified framework for differential and
difference equations, but does not by itself provide a continuous
interpolation of a given discrete solution. Impulsive differential equations
\citep{Lakshmikantham1989} describe continuous
evolution interrupted by instantaneous jumps, rather than constructing a
continuous trajectory from a given discrete solution. Conversely, the inverse
problem of discretisation, or ``continuization'' (the construction of a
continuous-time system from a discrete-time system), has been studied for linear
discrete systems, including through the use of the matrix logarithm
\citep{Verriest1991,Verriest1993}, while the question of whether a discrete
flow can be represented as the time-one map of a continuous flow has also
been studied as an embedding problem \citep{SharmaHicks1980}. 

Our approach is
complementary to these methods. Rather than discretising a differential
equation, relating only steady states, or restricting the construction to
linear systems, we start from an exact or approximate solution of a
discrete-time model and construct a differential equation whose solution
reproduces that solution at the discrete times. This provides a direct way
to compare the corresponding discrete-time and continuous-time descriptions.

\subsection{Terminology}

We note that while different examples may have very different interpretations
of the discrete time steps, in what follows we shall describe discrete time
steps as \textit{generations}, irrespective of the context. Additionally, when
the time is set equal to a generation time in a continuous-time solution, and
the solution \textit{precisely agrees} with the solution of the discrete-time
problem, we shall often describe the discrete-time and continuous-time problems as
being \textit{equivalent}.


\section{Elementary example: population growth}
\label{Elementary example section}

To focus our thoughts, let us consider the example of an extremely large
population of annual plants, whose numbers change deterministically.

Each plant produces seeds, some of which germinate and grow into mature
plants. If competition is fierce, such that every year each plant contributes,
on average, very close to one plant to the population in the following year,
then the population number exhibits small fractional (or percentage) changes
from year to year. The number of plants will obey a \textit{difference
equation}. This may be well approximated by the solution of a
\textit{differential equation}.

By contrast, if the plants are located in a sparsely populated area, with
reduced competition, then many of the seeds that each plant produces will grow
into mature plants. In such a situation, each plant contributes, on average,
appreciably more than one plant to the following year. The population number
will therefore exhibit large fractional changes from year to year and will
grow geometrically \citep{Crawley1990}. In such a case, it is a very bad
approximation to simply replace the discrete-time difference equation for the
population number by a continuous-time differential equation.

Here, such an example of deterministic population growth motivates our initial
considerations, where the number of plants obeys a difference equation, but
despite the fact that large fractional changes in the numbers may occur from
year to year, we will show that the number of plants may be calculated from
the solution of a differential equation. More precisely, the solution of the
differential equation, when evaluated at the discrete times, will coincide
with the solution of the difference equation.


\subsection{Description}

Let $x_{n}$ denote the size of a population in generation $n$ ($n=0,1,2,\ldots
$). The equation governing the size of the population is
\begin{equation}
x_{n+1}=\left(  1+r\right)  x_{n}. \label{simple}
\end{equation}
This describes the situation where the population size deterministically
changes by a factor $1+r$ each generation, with size treated as a continuous
non-negative quantity. Until we say otherwise, we take the parameter $r$ to be
a constant. If $r$ lies in the range $-1<r<0$ then the population size will
decrease each generation, while if $r>0$ then it will increase.

A full specification of the problem needs $x_{n}$ to be subject to an initial
condition. We take
\begin{equation}
x_{0}=a \label{x_0=a}
\end{equation}
where $a$ is the initial population size. The solution for $x_{n}$ is then
given by
\begin{equation}
x_{n}=(1+r)^{n} a\hspace{0.25cm}\text{with}\hspace{0.25cm}n=0,1,2,\ldots.
\label{soln difference}
\end{equation}

One possible way to obtain a continuous-time representation of Eq.
(\ref{simple}) -- an approximation that we will \textit{not} adopt -- is to write
the equation as $x_{n+1}-x_{n}=rx_{n}$. Then: (i) replace the discrete index,
$n$, by a continuous time, $t$; (ii) replace $x_{n}$ by the function $x(t)$;
(iii) replace $x_{n+1}-x_{n}$ by the derivative $dx(t)/dt$. This leads to the
differential equation $dx(t)/dt\simeq rx(t)$ with solution $x(t)=\exp(rt)a$.
The procedure outlined results in a reasonable approximation when $r$ is small
because the discrete-time and continuous-time problems have smooth solutions with
very similar shapes, i.e., replacing $n$ by $t$ in the factor $(1+r)^{n}$ in
Eq. (\ref{soln difference}) leads to $(1+r)^{t}=\exp(t\ln(1+r))=\exp
(rt)\times(1+O(tr^{2}))$. Thus when $|r|\ll1$ the quantities $(1+r)^{t}\ $and
$\exp(rt)$ differ appreciably only at very long times, of the order of
$r^{-2}$.

In this work, however, we do not take such a direct and limited approach. In
particular, given a discrete-time problem, with solution $x_{n}$, we look for
a continuous-time problem that has a solution $x(t)$, for $t$ continuous, such
that when $t$ is set equal to the integer value $n=0,1,2,...$ the solution to
the continuous-time problem \textit{precisely coincides} with the
corresponding value of the exact solution of the discrete-time problem. That
is $x(n)\equiv x(t=n)$ \textit{exactly} coincides with the discrete-time
solution, $x_{n}$, i.e.,
\begin{equation}
x(n)=x_{n}\hspace{0.25cm}\text{for}\hspace{0.25cm}n=0,1,2,....
\end{equation}


\subsection{Exact analysis, time-homogeneous case}

To proceed with an exact analysis of population growth, as defined by Eqs.
(\ref{simple}) and (\ref{x_0=a}), we begin with the exact solution in Eq.
(\ref{soln difference}), namely $x_{n}=(1+r)^{n}a$. In this, we replace the
integer-valued generation number, $n$, by the continuous-time parameter $t$
(with $t\geq0$). This leads us to define $x(t)=(1+r)^{t}a$, which we now view
as the solution of an unknown continuous-time problem.

At first sight, we seem to have gained little by this procedure. By
construction, when $t=0,1,2,...$ we have $x(t)$ precisely coinciding with the
values taken by the corresponding $x_{n}$. However, by differentiating
$x(t)=(1+r)^{t}a$ with respect to $t$, which is possible because $t$ is
continuous, we find
\begin{equation}
\frac{dx(t)}{dt}=\ln(1+r)x(t). \label{ode simple}
\end{equation}
In addition to this differential equation, the solution is subject to the same
initial condition as the discrete-time problem, i.e., $x_{0}=a$, which we now
write as
\begin{equation}
x(0)=a. \label{x(0)=a}
\end{equation}
With Eqs. (\ref{ode simple}) and (\ref{x(0)=a}) we have a fully specified
continuous-time problem.

It follows that with Eqs. (\ref{ode simple}) and (\ref{x(0)=a}) we have
established a continuous-time problem that represents a dynamical system with
the exact property that at the discrete times $t=0,1,2,\ldots$ the value of
the continuous-time solution precisely coincides with the solution of the 
discrete-time problem.

Since a finite difference is the discrete analogue of a derivative, we write
Eq. (\ref{simple}) in terms of a finite difference as $x_{n+1}-x_{n}=rx_{n}$,
and compare this equation with Eq. (\ref{ode simple}). We then see it is
possible to write down a \textit{mapping} between the parameters appearing in
the `equivalent' discrete-time and continuous-time problems. That is, the parameter
$r$ in the discrete-time equation (Eq. (\ref{simple})), which can be
interpreted as a `generational growth rate', becomes replaced by the
`instantaneous growth rate' $\ln(1+r)$ in the corresponding continuous-time
equation (Eq. (\ref{ode simple})). Thus we have the mapping
\begin{equation}
\begin{array}
[c]{ccc}
r & \longrightarrow & \ln(1+r)\\
\text{discrete time} &  & \text{continuous time.}
\end{array}
\label{elementary mapping}
\end{equation}
This is consistent in the sense that when $r$ vanishes there is no population
growth in both discrete and continuous time. Further, when $|r|$ is small
($|r|\ll1$) both $r$ and $\ln(1+r)$ are very similar in value and so
the mapping is not that informative. However, a feature of the above mapping
is that `equivalence' of the discrete-time and continuous-time problems does not
require the parameter $r$ to be small, and this has biological implications.

\begin{figure}[H]
\includegraphics[width=1\textwidth]{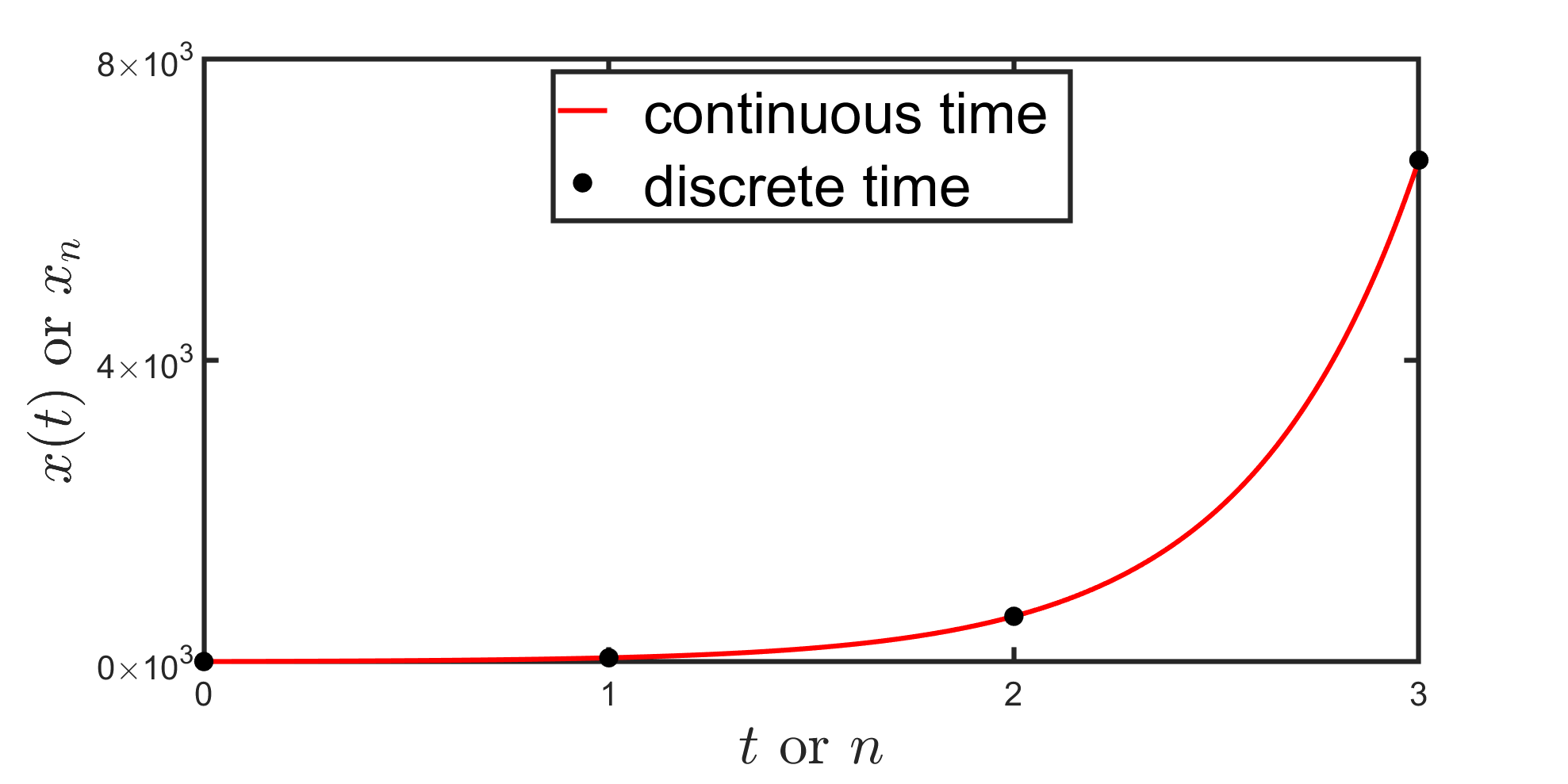} \centering
\caption{\textbf{Discrete and continuous solutions for elementary population
growth.} In this figure we plot the discrete-time and continuous-time solutions
associated with Eqs. (\ref{simple}) and (\ref{ode simple}), respectively. The
parameters adopted for the figure were $a=5$ and $r=10$.}
\label{fig:simple}
\end{figure}

As an example, setting $r=10$, in Eqs. (\ref{simple}) and (\ref{ode simple}),
which is definitely not a small value, does not stop the solution of Eq.
(\ref{ode simple}) from precisely coinciding, at $t=0,1,2,...$ with the
solution of Eq. (\ref{simple}), and in this case the `generational growth
rate' $r=10$ and the `instantaneous growth rate' $\ln(1+r)\equiv\ln
(11)\simeq2.4$ are very different.  Despite the extreme simplicity of this 
scenario, it has relevance to the
deterministic increase of a highly infectious disease that has a reproduction
number ($R_0$) of $11$, such as measles \citep{Guerra}. In continuous time,
the disease spread is described by dynamics with an instantaneous growth rate
$r = \ln(11) \simeq 2.4$. Under the fixed generation-interval assumption
implicit in the discrete model, this is precisely the standard relationship
between the growth rate and the reproduction number, $r = \ln R_0$
\citep{molles_ecology_2026}. The example thus illustrates this known
conversion but also shows that the continuous-time description precisely captures
an extremely rapid growth, as illustrated in Figure \ref{fig:simple}.


As illustrated in Figure \ref{fig:simple}, at integer times, the 
continuous-time solution, that arises from an ordinary differential equation, precisely
coincides with the discrete-time solution, despite the fact that the solution
becomes large (in the thousands) in a very few time steps.

In the above example, as partially summarised by Eq. (\ref{elementary mapping}
), we have established a mapping of parameters between a problem with discrete
time and one with continuous time. We next investigate how the procedure works
when the parameter $r$ has time dependence.


\subsection{Exact analysis, time-inhomogeneous case}

\label{Exact pop, inhomog}

Time inhomogeneity plays a key role in many biological problems. As an
example, transitions of epidemics seen over the years have been explained in
terms of \textquotedblleft an exogenous factor -- slow variation in the
average rate of recruitment of new susceptibles\textquotedblright
\ \cite{earn2000}.

Here, we consider the case where the parameter $r$ in the population growth
problem has \textit{time dependence}. In particular, we now take the
generalisation of Eq. (\ref{simple}) to be
\begin{equation}
x_{n+1}=\left[  1+r(n+1)\right]  x_{n}\hspace{0.25cm}\text{with}
\hspace{0.25cm}x_{0}=a. \label{simple inhomog}
\end{equation}
The solution is
\begin{align}
x_{n}  &  =[1+r(n)][1+r(n-1)]...[1+r(1)] a\nonumber\\
& \nonumber\\
&  =\exp\left(  \sum_{k=1}^{n}\ln(1+r(k))\right)  a.
\label{exact soln simple inhomog}
\end{align}
Simply replacing $n$ by $t$ in Eq. (\ref{exact soln simple inhomog}) does not
obviously yield a meaningful continuous-time analogue of $x_{n}$. However
there are many forms of $\ln(1+r(k))$, and hence of $r(k)$, where we can
evaluate the sum in Eq. (\ref{exact soln simple inhomog}) in closed-form, and
from this find an exact expression for $x_{n}$. For such forms of $r(k)$ we
can obtain the \textit{equivalent} continuous-time problem.


\subsubsection{Particular example.}

As a particular example, consider
\begin{equation}
\ln(1+r(k))=b+ck\text{ or equivalently }r(k)=\exp(b+ck)-1 \label{r special}
\end{equation}
with $b$ and $c$ constants. Since there is a closed-form expression for
$\sum_{k=1}^{n}k$, namely $n(n+1)/2$, we arrive at the explicit expression
$x_{n}=\exp\left(  bn+cn(n+1)/2\right)  a$. Simply replacing $n$ by $t$ in
this expression yields
\begin{equation}
x(t)=\exp\left(  bt+c\frac{t(t+1)}{2}\right)  a. \label{special cont t soln}
\end{equation}
This form of $x(t)$ obeys $dx(t)/dt=\left(  b+c\left(  t+\tfrac{1}{2}\right)
\right)  x(t)$ with $x(0)=a$, which we can write as
\begin{equation}
\frac{dx(t)}{dt}=\ln\big(1+r\left(  t+\tfrac{1}{2}\right)  \big)x(t)\hspace
{0.25cm}\text{with}\hspace{0.25cm}x(0)=a. \label{p+qt ode}
\end{equation}
Thus Eq. (\ref{exact soln simple inhomog}), with $r(k)=\exp\left(
b+ck\right)  -1$, has an exact solution given by $x_{n}=x(n)$, with $x(t)$ the
solution of Eq. (\ref{p+qt ode}), which describes continuous time behaviour.

To determine the exact mapping between discrete and continuous problems, we
rewrite Eq. (\ref{simple inhomog}) as $x_{n+1}-x_{n}=r(n+1)x_{n}$. Comparing
this with Eq. (\ref{p+qt ode}) we find a more complicated mapping than in the
time-homogeneous case, namely
\begin{equation}
\begin{array}
[c]{ccc}
r(n+1) & \longrightarrow & \ln\left(  1+r\left(  t+\tfrac{1}{2}\right)  \right) \\
\text{discrete time} &  & \text{continuous time.}
\end{array}
\label{special mapping}
\end{equation}
This is intuitively plausible; the relevant time-dependent growth rate in the
continuous-time problem is evaluated at half a time step away from that in the
discrete problem.


\subsubsection{Illustration of the particular example}

\begin{figure}[H]
\includegraphics[width=1\textwidth]{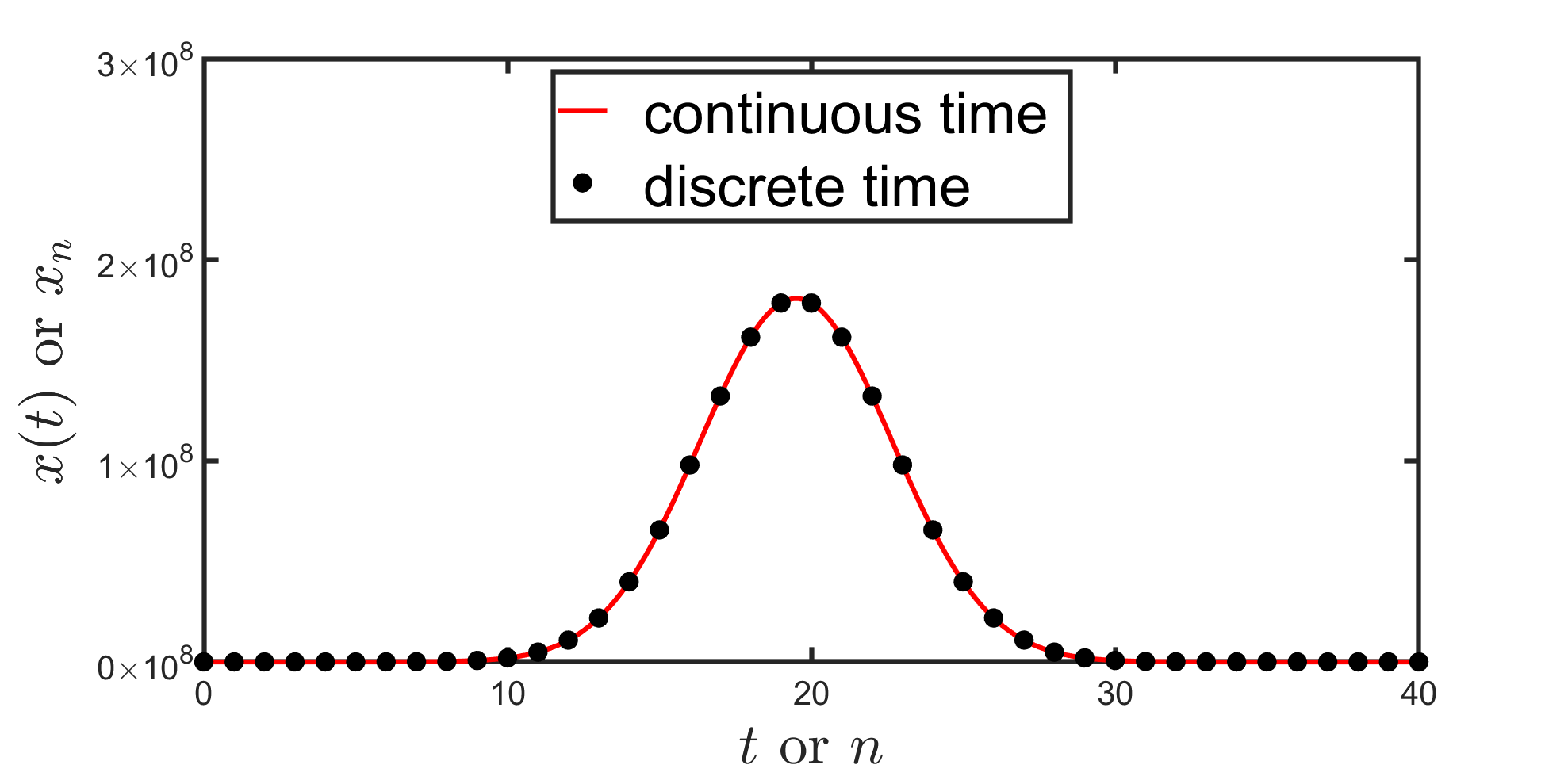} \centering
\caption{\textbf{Discrete and continuous solutions for an exactly soluble
form of $\bm{r(k)}$.} In this figure we plot the discrete and continuous
solutions associated with Eqs. (\ref{simple inhomog}) and (\ref{p+qt ode}),
respectively, for the special, exactly soluble form of $r(k) $ given in Eq.
(\ref{r special}), namely $r(k)=\exp{(b+ck)}-1$. The parameters adopted for
the figure were $a=1$, $b=2$, and $c=-1/10$.}
\label{fig:simple inhomog}
\end{figure}

The exactly soluble case $\ln(1+r(k))=b+ck$ of Eq. (\ref{r special}), despite
its simplicity, is not devoid of biologically interesting behaviour. To
illustrate this we show, in Figure \ref{fig:simple inhomog}, the equivalent
continuous-time and discrete-time solutions, $x(t)$ and $x_{n}$, respectively, for
a range of $t$ or $n$, for a particular choice of $a$, $b$ and $c$.

\bigskip


The parameter choices adopted for Figure \ref{fig:simple inhomog} represent a
model of an environment that degrades over time (or the deterministic spread
of a disease in a population with a decreasing reproduction number). For early
times ($k<20$) the value of $r(k)$ used in Figure \ref{fig:simple inhomog} (namely $r(k)=\exp
{(2-k/10)}-1$) is positive, and the population is growing, but for later times
($k>20$) the value of $r(k)$ is negative and the population is decreasing.
This behaviour is exhibited in Figure \ref{fig:simple inhomog}.

A striking feature of Figure \ref{fig:simple inhomog} is that from a very
modest initial population size ($a=1$), \textit{very} large values of the
population size ($\sim10^{8}$) are rapidly achieved (i.e., in $20$
generations), yet the continuous-time solution fully matches this behaviour at
the positive integral times.

In Appendix A we numerically illustrate the above relation between the discrete-time
and continuous-time problems, as captured by Eqs. (\ref{simple inhomog}) and
(\ref{p+qt ode}). The results of Appendix A give an indication that the sizes
of the errors arising from machine precision (or aspects of numerical
evaluation) are of the order of $1$ part in $10^{14}$.

\subsection{Approximate analysis, time-inhomogeneous case}

\label{Approx pop, inhomog}

We can exploit the results established so far, and carry out what is generally
an approximate analysis of Eq. (\ref{simple inhomog}). We start with the exact
solution in Eq. (\ref{exact soln simple inhomog}). This equation involves the
quantity $\ln(1+r(k))$. Using the elementary `mid-point' integration rule
\cite{Sauer2012}
\begin{equation}
\int_{k-1}^{k}h\left(  q\right)  dq\simeq h\left(  k-\frac{1}{2}\right)
\label{mid-point rule}
\end{equation}
but in the \textit{reverse direction}, allows us to approximate $\ln(1+r(k))$
by an integral:
\begin{equation}
\ln(1+r(k))\simeq\int_{k-1}^{k}\ln\left(  1+r\left(  q+\tfrac{1}{2}\right)
\right)  dq. \label{mid-point}
\end{equation}
Then Eq. (\ref{exact soln simple inhomog}) has the approximation
\newline$x_{n}\simeq\exp\left(  \sum_{k=1}^{n}\int_{k-1}^{k}\ln\left(
1+r\left(  q+\tfrac{1}{2}\right)  \right)  dq\right)  a$ and combining all of
the $q$ integrals yields $x_{n}\simeq\exp\left(  \int_{0}^{n}\ln\left(
1+r\left(  q+\tfrac{1}{2}\right)  \right)  dq\right)  a$. Reserving the use of
$x(t)$ for the continuous function of $t$ that is \textit{exactly} equivalent
to $x_{n}$, we shall use $x^{(\mathrm{app})}(t)$ for the \textit{approximately
} equivalent solution. In the present case, we thus have
\begin{equation}
x^{(\mathrm{app})}(t)=\exp\left(  \int_{0}^{t}\ln\left(  1+r\left(
q+\tfrac{1}{2}\right)  \right)  dq\right)  a \label{xapp}
\end{equation}
which is a function of continuous time, $t$, but it is not, generally, exactly
equivalent to $x_{n}$. We find, from Eq. (\ref{xapp}), that $x^{(\mathrm{app}
)}(t)$ obeys
\begin{equation}
\frac{dx^{(\mathrm{app})}(t)}{dt}=\ln\big(1+r\left(  t+\tfrac{1}{2}\right)
\big)x^{(\mathrm{app})}(t) \label{ode simple not homog}
\end{equation}
with $x^{(\mathrm{app})}(0)=a$. Thus $x_{n}$, which satisfies Eq.
(\ref{simple inhomog}), has the approximation
\begin{equation}
x_{n}\simeq x^{(\mathrm{app})}(n).
\end{equation}

Plausibly, when the function $r(t)$ changes slowly with $t$, we have that Eq.
(\ref{ode simple not homog}) is an approximate continuous-time analogue of Eq.
(\ref{simple inhomog}) and we have the \textit{generally approximate mapping}
\begin{equation}
\begin{array}
[c]{ccc}
r(n+1) & \longrightarrow & \ln\left(  1+r\left(  t+\tfrac{1}{2}\right)  \right)
.\\
\text{discrete time} &  & \text{continuous time}
\end{array}
\label{approximate mapping}
\end{equation}

It seems likely that the solution of Eq. (\ref{ode simple not homog}) has the
ability to capture a lot of the behaviour of the exact solution even for
values of $r$ that may be large, since for constant $r$, Eq.
(\ref{ode simple not homog}) yields the \textit{exact} continuous-time
analogue of Eq. (\ref{simple inhomog}), even for very large $r$. However,
matters are slightly better than this. The mid-point integration rule in Eq.
(\ref{mid-point rule}) works without error if the function in this equation,
$h(q)$, is a linear function, i.e., when $h(q)=b+cq$. Since we applied the
mid-point rule to the function $\ln(1+r(k))$, this means when $\ln(1+r(k))$ is
a linear function of $k$ the mid-point rule will be exact. Thus when
$\ln(1+r(k))=b+ck $ the solution of Eq. (\ref{ode simple not homog}) will
precisely reproduce the discrete-time solution at $t=0,1,2,\ldots$. We have
already seen the exactness of this in the previous section, where $r(k)$ was
given by Eq. (\ref{r special}). We can thus say, for $\ln(1+r(k))$ a linear
function of $k$, that $x^{(\mathrm{app})}(t)$ is not an approximation but is
exact, i.e., $x^{(\mathrm{app})}(t)=x(t)$.

Generally, $\ln(1+r(k))$ is not a linear function of $k$ and the use of the
midpoint integration rule in Eq. (\ref{exact soln simple inhomog}) leads to an
error in Eq. (\ref{xapp}). The error is proportional to the second derivative of
$\ln(1+r(k))$ \cite{Sauer2012} and using this, we can obtain an analytical
bound on the magnitude of the error.


\subsubsection{Illustrating accuracy with a specific nonlinear form of
$\bm{r(t)}$}

We now consider an example of an $r(k)$ that illustrates the working/accuracy
of Eq. (\ref{ode simple not homog}) when $\ln\left(  1+r\left(  k\right)
\right)  $ is a \textit{nonlinear} function of $k$, so $x^{(\mathrm{app} )}(t)
$ is a genuine approximation of the equivalent continuous-time solution. We
choose a nonlinear form of $r(k)$ given by
\begin{equation}
r(k)=\exp\left(  3bc\sin\left(  c\left(  k-\tfrac{1}{2}\right)  \right)
\cos^{2}\left(  c\left(  k-\tfrac{1}{2}\right)  \right)  \right)  -1
\label{specific r}
\end{equation}
where $b$ and $c$ are independent constants. We have adopted this form of
$r(k)$ for illustration because it leads to a simple form of the approximate
equivalent solution, $x^{(\mathrm{app})}(t)$, where the parameter $b$
primarily determines the magnitude of the solution, while the parameter $c$
controls its periodicity (see Eq. (\ref{solution simple r(t) continuous})).

\begin{figure}[H]
\includegraphics[width=1\textwidth]{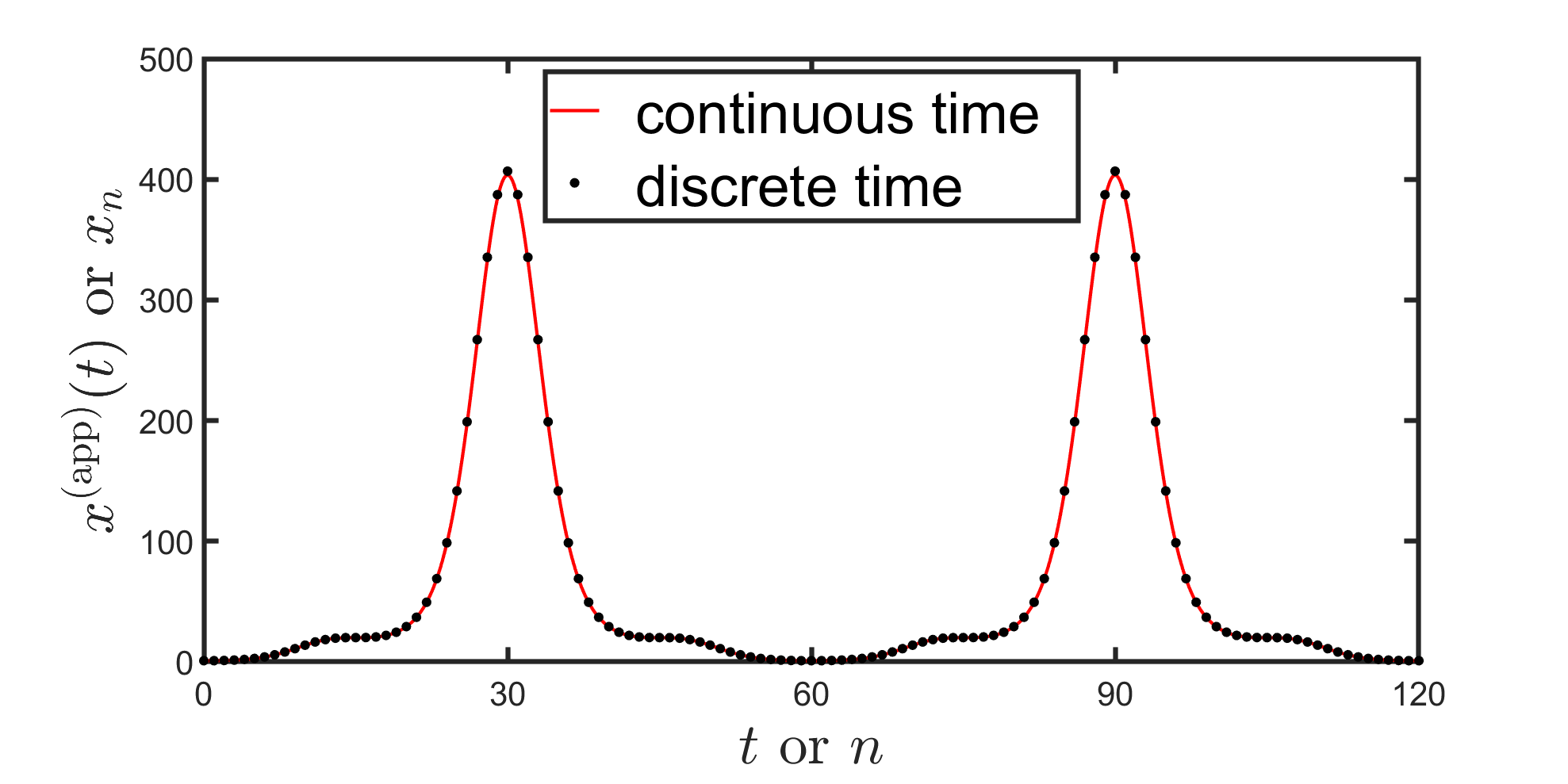} \centering
\caption{\textbf{Discrete and approximate continuous solutions for a
particular nonlinear form of $\bm{r(k)}$.} In this figure we plot the discrete
and approximate equivalent solutions for the specific form of $r(k)$ given in
Eq. (\ref{specific r}). The discrete-time solution, $x_{n}$, was obtained from
Eq. (\ref{exact soln simple inhomog}) by iteration, while the approximate
equivalent solution, $x^{(\mathrm{app})}(t)$, is analytically given in Eq.
(\ref{solution simple r(t) continuous}). The parameters adopted for the figure
were $a=1$, $b=3$, and $c=2\pi/60 \simeq0.1047$.}
\label{fig: simple approx inhomog}
\end{figure}

For the above form for $r(k)$, Eq. (\ref{ode simple not homog}) yields\newline
$dx^{(\mathrm{app})}(t)/dt=3bc\sin\left(  ct\right)  \cos^{2}\left(
ct\right)  x^{(\mathrm{app})}(t)$, subject to $x^{(\mathrm{app})}
(0)=a$. The solution is
\begin{align}
x^{(\mathrm{app})}(t)  &  =\exp\left(  \int_{0}^{t}3bc\sin(cq)\cos
^{2}(cq)dq\right)  a\nonumber\\
& \nonumber\\
&  =\exp\left(  b(1-\cos^{3}\left(  {ct}\right)  )\right)  a.
\label{solution simple r(t) continuous}
\end{align}
The solution in this example behaves periodically over time. To illustrate the
behaviour, we show, in Figure \ref{fig: simple approx inhomog}, the form of
the approximate equivalent solution, $x^{(\mathrm{app})}(t)$, and the 
discrete-time solution, $x_{n}$, that was obtained by iteration of Eq.
(\ref{simple inhomog}).


From Figure \ref{fig: simple approx inhomog} there appears to be very good
agreement between the continuous-time and discrete-time solutions, even though the
solutions exhibit nontrivial variation, with the height of the functions
ranging from $\sim0$ to $\sim400$.

\begin{figure}[H]
\includegraphics[width=1\textwidth]{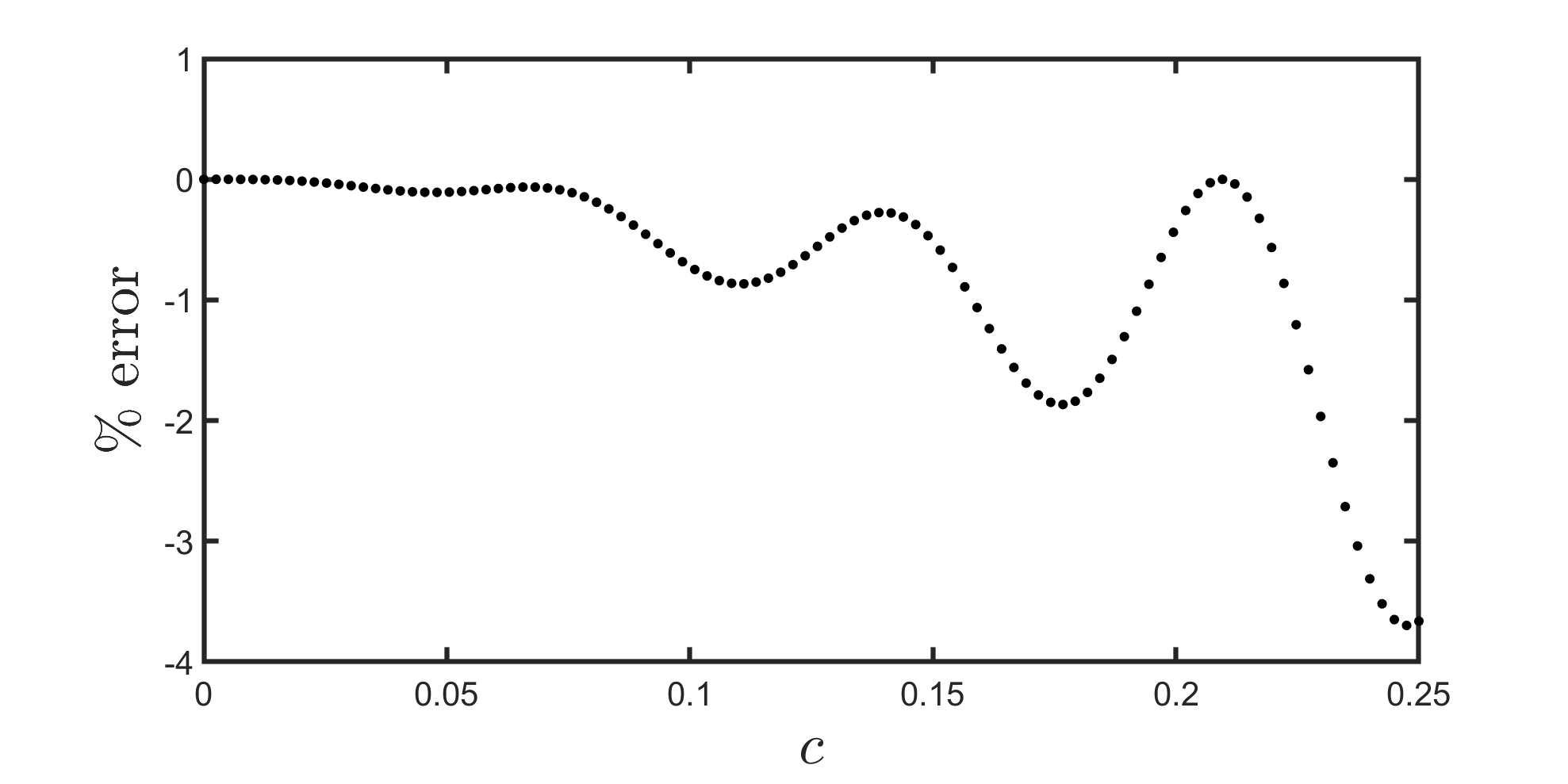} \centering
\caption{\textbf{Error of approximate continuous solutions for a particular
form of $\bm{r(k)}$.} In this figure we plot the percentage error of the
approximate continuous solution plotted in Figure 3, when evaluated at
$t=n=30$. The plot applies for the specific form of $r(k)$ given in Eq.
(\ref{specific r}). The figure was obtained by setting $t=30$ in the
approximate continuous-time solution, $x^{(\mathrm{app})}(t)$ of Eq.
(\ref{solution simple r(t) continuous}) and then, for a range of $c$ values,
numerically computing the percentage error $[x^{(\mathrm{app}
)}(30)-x_{30}]/x_{30} \times100$, with $x_{30}$ obtained from iteration of Eq.
(\ref{simple inhomog}). The values of the parameters $a$ and $b$ were those
used in Figure 3.}
\label{fig:simple approx inhomog error}
\end{figure}

For the values of $a$ and $b$ that were used in Figure
\ref{fig: simple approx inhomog} we have numerically compared the error
between $x^{(\mathrm{app})}(t=n)$ and $x_{n}$, for a range of $c$ values, with
the results given in Figure \ref{fig:simple approx inhomog error}.


It is evident from Figure \ref{fig:simple approx inhomog error} that the
errors are $\sim$ twelve orders of magnitude larger than those typical of
numerical evaluation (see Appendix A) and thus represent genuine errors of the
method. Furthermore, the errors have a tendency to increase with the value of
the parameter $c$, that influences the rate of change of the solution (see Eq.
(\ref{solution simple r(t) continuous} )). This is understandable: the
approximate continuous-time solution does not precisely reproduce the 
discrete-time solution when there are relatively rapid changes in $r(k)$. However, we
see from Figure \ref{fig:simple approx inhomog error} that for the $c$ value
adopted for Figure 3 (namely $c\simeq0.1047$), the errors in the approximate
continuous-time solution are relatively small (the errors have a magnitude
less than $0.9\%$).


\section{A nonlinear problem from genetics}

\label{Genetics section}

In this section we illustrate the above procedure, of going from discrete to
continuous time, in a model from genetics.


\subsection{Description}

We consider a very large population of asexual organisms that carry one gene,
which can be either the $A$ or $B$ allele. For simplicity, we shall often
refer to the type of such organisms as $A$ and $B$. The population has
discrete (or non-overlapping) generations, labelled by $n=0,1,2,\ldots$. The
census point in a generation is taken to be the adult stage, which occurs at
the beginning of a generation, immediately prior to reproduction.

\noindent The life cycle is:

\noindent(i) adults produce offspring, which may contain a mutation, in which
case the gene of the offspring will differ from that of the parent;

\noindent(ii) the adults die after reproduction;

\noindent(iii) the offspring population undergoes number regulation, with $N$
survivors constituting the adults of the next generation.\medskip

For simplicity, we assume that each carrier of an $A$ or $B$ allele produces a
total of $\alpha$ or $\beta$ offspring, respectively. Thus $\alpha$ and
$\beta$ represent fertilities and, until we say otherwise, $\alpha$ and
$\beta$ are \textit{constants}.

We shall focus on the relative frequencies (or proportions) of the population
that carry the different genes. Because we deal with frequencies, all results
depend on the \textit{ratio} of $\alpha$ and $\beta$ (see Appendix B) and we
define a \textit{selection coefficient}, $s$, by
\begin{equation}
\frac{\alpha}{\beta}=1+s. \label{s def}
\end{equation}
The value of $s$ is a measure of the strength of selection that is acting on
$A$ relative to $B$ and $s$ lies in the range $-1\leq s<\infty$. While
selection coefficients in nature are often small ($|s|\ll1$) \cite{Eyrewalker}
there are known cases of large $|s|$. As an example, if $s$ is in the vicinity
of $-1$, then it corresponds to near lethality of the $A$ gene
\cite{WaxmanOverall2020}. To leave the analysis general, we make no
assumptions about the value of $s$.

Mutations are assumed to occur independently in the production of each
offspring. We use $u$ to denote the probability of an $A$ allele in a parent
mutating to a $B$ allele in an offspring, and $v$ is the corresponding $B$ to
$A$ mutation probability.

We assume a number-regulating mechanism reduces the population size, by
randomly picking $N$ individuals, who constitute the adults at the beginning
of the next generation. Assuming $N$ is very large, number regulation
negligibly affects gene frequencies and we can treat the dynamics as
\textit{deterministic}.


\subsection{Exact analysis, time-homogeneous case}

\label{Exact analysis genetics section}

Let $x_{n}$ represent the relative frequency (or proportion) of adults that
carry the $A$ gene in generation $n$, with $1-x_{n}$ the corresponding
frequency of adult carriers of the $B$ gene. In Appendix B we show that
$x_{n}$ obeys the equation
\begin{equation}
x_{n+1}=x_{n}+\sigma(x_{n})x_{n}(1-x_{n})-U(x_{n})x_{n}+V(x_{n})(1-x_{n})
\label{x n+1 genetics}
\end{equation}
where
\begin{equation}
\sigma(x)=\frac{s}{1+sx},\qquad U(x)=\frac{u(1+s)}{1+sx},\qquad V(x)=\frac
{v}{1+sx}. \label{sigma u and v}
\end{equation}

We shall proceed with Eq. (\ref{x n+1 genetics}), without making assumptions
about parameter values, in particular, allowing $s$ to be appreciable as it
can be in some cases \citep{WaxmanOverall2020}.

Equation (\ref{x n+1 genetics}) is a nonlinear difference equation for
$x_{n}$ that can be analytically solved. In Appendix C we show that with
initial frequency $x_{0}=a$, we have
\begin{equation}
x_{n}=\hat{x}+\frac{1}{\frac{\lambda^{n}}{a-\hat{x}}+\frac{1-\lambda^{n}
}{1-\lambda}\mu} \label{genetics soln}
\end{equation}
where
\begin{align}
\hat{x}=  &  \frac{1}{2s}\Bigg\{(1-u)(1+s)-(1+v)\Bigg.\nonumber\\
& \nonumber\\
&  +\Bigg.\sqrt{\left[  (1-u)(1+s)-(1-v)\right]  ^{2}+4uv(1+s)}\Bigg\},\\
& \nonumber\\
\lambda &  =\frac{(1+s\hat{x})^{2}}{(1+s)(1-u-v)}\quad\text{and}\quad\mu
=\frac{s(1+s\hat{x})}{(1+s)(1-u-v)}.
\end{align}

We now replace the integer-valued generation number, $n$, in Eq.
(\ref{genetics soln}) by the continuous time parameter, $t$, leading us to
\begin{equation}
x(t)=\hat{x}+\frac{1}{\frac{\lambda^{t}}{a-\hat{x}}+\frac{1-\lambda^{t}
}{1-\lambda}\mu}.
\end{equation}
As before, we view $x(t)$ as the solution of an unknown continuous-time
problem. On differentiating this form of $x(t)$ with respect to $t$, and then
eliminating all dependence on the initial frequency, $a$, we obtain the 
time-homogeneous problem
\begin{equation}
\frac{dx(t)}{dt}=\sigma_{c}x(1-x)-U_{c}x+V_{c}(1-x)\hspace{0.25cm}
\text{where}\hspace{0.25cm}x(0)=a \label{continue}
\end{equation}
and
\begin{align}
\sigma_{c}  &  =-\mu\frac{\ln{\lambda}}{1-\lambda},\\
& \nonumber\\
U_{c}  &  =(1-\hat{x})\left[  1-\lambda-\mu(1-\hat{x})\right]  \frac
{\ln{\lambda}}{1-\lambda},\\
& \nonumber\\
V_{c}  &  =\hat{x}\left(  1-\lambda+\mu\hat{x}\right)  \frac{\ln{\lambda}
}{1-\lambda}.
\end{align}
We note that the coefficients of $x(1-x)$, $x$ and $1-x$, in Eq.
(\ref{continue}) are all constants.

Comparing Eq. (\ref{continue}) with
Eq. (\ref{x n+1 genetics}), written in the form $x_{n+1}-x_{n}=\sigma
(x_{n})x_{n}(1-x_{n})-U(x_{n})x_{n}+V(x_{n})(1-x_{n})$, we infer the exact
mappings, from \textit{frequency-dependent} coefficients in discrete time, to
\textit{frequency independent} coefficients in continuous time. These are
\begin{equation}
\begin{array}
[c]{ccc}
\sigma(x)=\frac{s}{1+sx}\text{ } & \longrightarrow & \sigma_{c}\text{ (independent
of }x\text{)}\\
&  & \\
U(x)=\frac{u(1+s)}{1+sx} & \longrightarrow & U_{c}\text{ (independent of
}x\text{)}\\
&  & \\
V(x)=\frac{v}{1+sx} & \longrightarrow & V_{c}\text{ (independent of }x\text{)}\\
&  & \\
\text{discrete time} &  & \text{continuous time.}
\end{array}
\label{genetics mapping}
\end{equation}

\subsubsection{Special case}

In the special case $u=v=0$, the coefficient of $x_{n}(1-x_{n})$ in Eq.
(\ref{x n+1 genetics}) and the coefficient of $x(1-x)$ in Eq. (\ref{continue})
are non-zero and these play an equivalent role in the dynamics. The
corresponding mapping is
\begin{equation}
\left.
\begin{array}
[c]{ccc}
\sigma(x)=\dfrac{s}{1+sx} & \longrightarrow & \sigma_{c}=\ln(1+s)\\
\text{discrete time} &  & \text{continuous time.}
\end{array}
\right\}  \quad\text{ special case}\label{genetics special}
\end{equation}
 which is a variant of the standard Wright/Malthusian conversion, see Section~1.2, Eq.~1.2.3 of \cite{CrowKimura1970}.

The mapping between coefficients in Eq. (\ref{genetics special}) has a
non-trivial character. It is not a mapping between parameters, but rather, a
mapping between \textit{functions} with different $x$ dependence. For example,
for strong negative selection ($s\sim-1$), when $x$ ranges from $0$ to $1$ the
coefficient $\sigma(x)$ varies from $-|s|$ to $-|s|/(1-|s|)$ which can be
considerable, while $\sigma_{c}$ has the value $\ln(1-|s|)$ which may be
large, but is constant. 


\subsection{Exact analysis, time-inhomogeneous case}

\label{Exact analysis inhomog genetics section}

We now consider a time-inhomogeneous genetics problem, where there is no
mutation but time-dependent selection acts. In this case we take the analogue
of Eq. (\ref{x n+1 genetics}) to be
\begin{equation}
x_{n+1}=x_{n}+\frac{s(n+1)}{1+s(n+1)x_{n}}x_{n}(1-x_{n})
\label{xn+1 genetics inhomog}
\end{equation}
where, now, the selection coefficient depends on the discrete time, $n$. In
Appendix D we show that the exact solution to Eq. (\ref{xn+1 genetics inhomog}
), subject to $x_{0}=a$, is
\begin{equation}
x_{n}=\left\{
\begin{array}
[c]{ll}
a, & n=0\\
& \\
\frac{a}{\left(  1-a\right)  \exp\left(  -\sum_{k=0}^{n-1}\ln\left(
1+s(k+1)\right)  \right)  +a}, & n\geq1.
\end{array}
\right.  \label{exact genetics inhomog}
\end{equation}
There are similarities of this result with the results for population growth
in Sections \ref{Exact pop, inhomog} and \ref{Approx pop, inhomog}. We could
replicate the sort of things we did in those sections, for example, choosing
specific forms for $\ln\left(  1+s(k+1)\right)  $ where the sum in Eq.
(\ref{exact genetics inhomog}) can be evaluated in closed-form, but the
similarity is sufficiently close that we will not do so here. We will,
however, approximate $\ln\left(  1+s\left(  k+1\right)  \right)  $ by an
\textit{integral} (again using the approximate mid-point integration rule in
the opposite direction):
\begin{equation}
\ln\left(  1+s\left(  k+1\right)  \right)  \simeq\int_{k}^{k+1}\ln\left(
1+s\left(  q+\tfrac{1}{2}\right)  \right)  dq. \label{approx}
\end{equation}
This result allows us to write Eq. (\ref{exact genetics inhomog}) for
$n=0,1,2,...$ as
\begin{equation}
x_{n}\simeq\frac{a}{\left(  1-a\right)  \exp\left(  -\int_{0}^{n}\ln\left(
1+s\left(  q+\frac{1}{2}\right)  \right)  dq\right)  +a}.
\label{genetic approx}
\end{equation}
Equation (\ref{genetic approx}) motivates defining the approximate,
equivalent, continuous-time solution
\begin{equation}
x^{(\mathrm{app})}(t)=\frac{a}{\left(  1-a\right)  \exp\left(  -\int_{0}
^{t}\ln\left(  1+s\left(  q+\frac{1}{2}\right)  \right)  dq\right)  +a}
\end{equation}
which satisfies
\begin{equation}
\frac{dx^{(\mathrm{app})}(t)}{dt}=\ln\left(  1+s\left(  t+\tfrac{1}{2}\right)
\right)  \times x^{(\mathrm{app})}(t)\left[  1-x^{(\mathrm{app})}(t)\right]  .
\label{ode special}
\end{equation}
Comparing Eq. (\ref{ode special}) with the original difference equation, Eq.
(\ref{xn+1 genetics inhomog}), written as $x_{n+1}-x_{n}=s(n+1)/(1+s(n+1)x_{n}
)x_{n}(1-x_{n})$, leads to the (generally approximate) mapping
\begin{equation}
\begin{array}
[c]{ccc}
\dfrac{s(n+1)}{1+s(n+1)x_{n}} & \longrightarrow & \ln\left(  1+s\left(
t+\tfrac{1}{2}\right)  \right) \\
\text{discrete time} &  & \text{continuous time.}
\end{array}
\end{equation}
As in the constant $s$ case of Section \ref{Exact analysis genetics section},
a frequency-dependent coefficient (namely that of $x_{n}(1-x_{n})$) in the
discrete-time problem becomes a frequency independent coefficient in
continuous time. Furthermore, the approximation in Eq. (\ref{approx}) is
\textit{exact} when $\ln\left(  1+s\left(  t+\frac{1}{2}\right)  \right)  $ is
a linear function of $t$. In this case, Eq. (\ref{ode special}) will have a
solution that precisely agrees with the solution of Eq.
(\ref{xn+1 genetics inhomog}) when $t=n$, irrespective of the size of the
parameters in $s(t)$.


\section{Oscillatory $x_{n}$ and complex $x(t)$}

It seems plausible that the method we have presented will also work for 
second-order difference equations, such as the Fibonacci problem (see e.g.,
\cite{Elaydi}), which represents a population where the number of individuals
at time $n$, written $x_{n}$, takes integer values and obeys $x_{n+2}
=x_{n+1}+x_{n}$ along with $x_{0}=0$ and $x_{1}=1$. The solution is
$x_{n}=c\left[  R^{n}-(-1/R)^{n}\right]  $ where $c=1/\sqrt{5}$ and
$R=(1+\sqrt{5})/2$. Note that in this solution the term involving $(-1/R)^{n}$
is oscillatory: it changes sign as $n\rightarrow n+1$. Thus despite the fact
that for $n>1$ the magnitude of $(-1/R)^{n}$ is smaller than the $R^{n}$ term,
possibly much smaller, the oscillatory behaviour of $(-1/R)^{n}$ is a new
feature that we have not so far encountered.

Indeed when we replace $n$ by $t$, the continuous-time solution $x(t)=c\left[
R^{t}-(-1/R)^{t}\right]  $ represents an ambiguously specified complex
solution, because of the presence of $(-1/R)^{t}$. We thus have to understand
oscillatory problems in discrete time, and their continuous time equivalent
analogues. We next study the simplest problem with this feature.


\subsection{Simplest example of an oscillatory solution}

\label{Simple Osc Section}

\begin{figure}[H]
\includegraphics[width=1\textwidth]{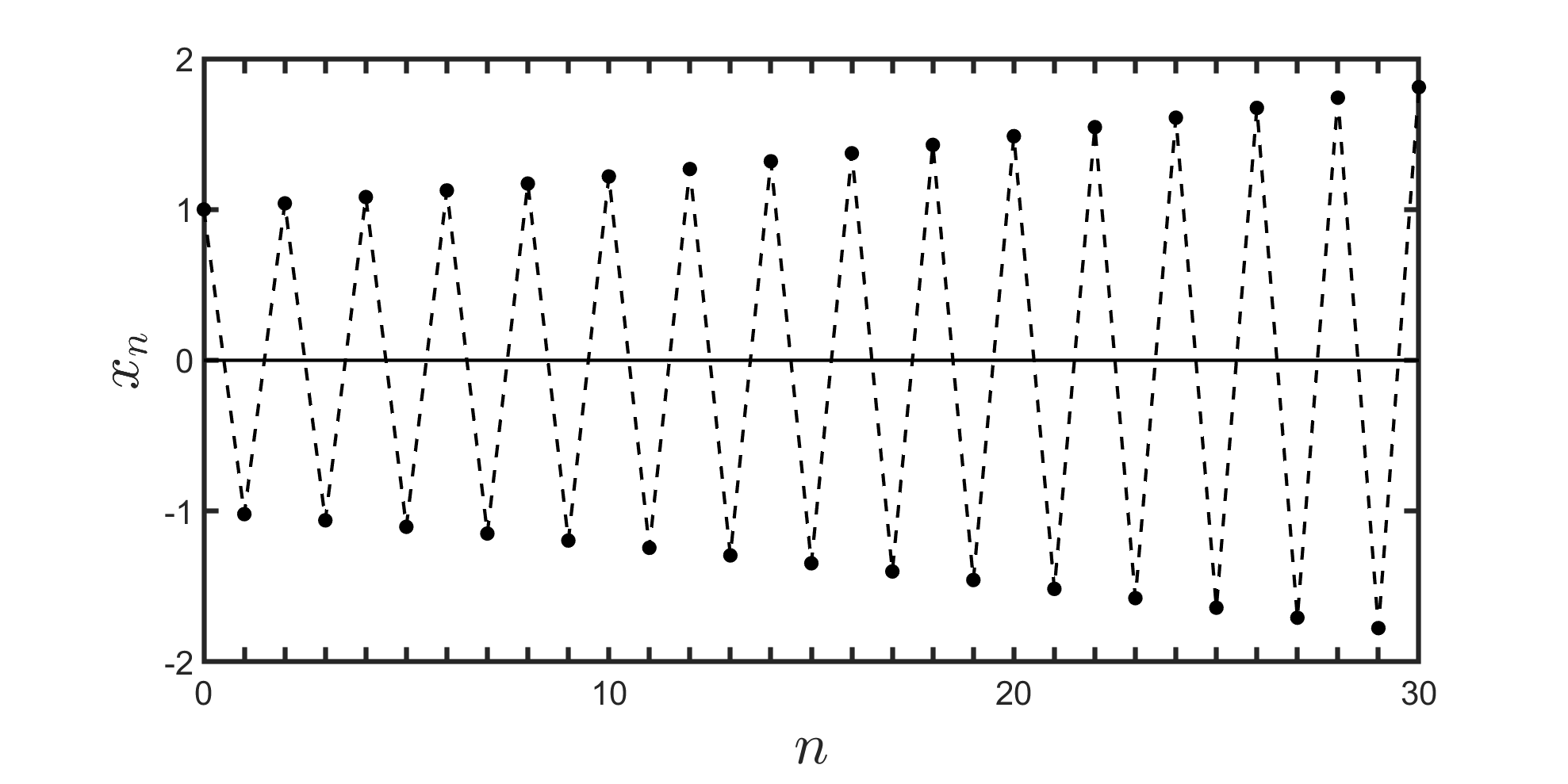} \centering
\caption{\textbf{Plot of an oscillatory} ${\bm x_{n}}$ \textbf{against}
{$\bm n$.} In this figure we plot $x_{n}=(-m)^{n}$ against $n$ as black dots.
To aid visualisation, we have joined the dots by dashed lines. The value of
the parameter $m$ used was $1.02$.}
\label{fig:simple oscillatory}
\end{figure}

Consider the discrete-time problem
\begin{equation}
x_{n+1}=zx_{n}\hspace{0.25cm}\text{with}\hspace{0.25cm}x_{0}=a \label{x'=Lx}
\end{equation}
whose solution is
\begin{equation}
x_{n}=z^{n}a. \label{x=L^n a}
\end{equation}
Thus if we take $z$ to be negative, say $z=-m$ with $m>0$, then the solution
for $x_{n}$ is
\begin{equation}
x_{n}=(-m)^{n}a\hspace{0.25cm}\text{ (the solution for }z=-m)
\label{subs l=-m}
\end{equation}
and this is oscillatory in the sense that $x_{n}$ and $x_{n+1}$ have different
signs. We have plotted an example of $x_{n}$ against $n$ in Figure
\ref{fig:simple oscillatory}.


Consider the solution of the continuous-time problem that is
\textit{equivalent} to Eq. (\ref{x'=Lx}), and obtained from Eq. (\ref{x=L^n a}
) by replacing $n$ by $t$:
\begin{equation}
x(t)=z^{t}a. \label{equiv x'=Lx}
\end{equation}
This satisfies
\begin{equation}
\frac{dx(t)}{dt}=\ln(z)x(t)\hspace{0.25cm}\text{with}\hspace{0.25cm} x(0)=a.
\label{equiv x'=Lx ode}
\end{equation}
To determine what arises when $z\rightarrow-m$, we cannot simply set $z=-m$
since this leaves the value of the logarithm ambiguous. A way to proceed is to
treat $z$ as a \textit{complex variable}. We then need to ensure that both
$z^{t}$ and $\ln(z)$ are well-defined functions of $z$ for general (real) $t$.
This entails both of these functions having a \textit{cut} along the negative
real axis in the complex $z$ plane. The way to arrive at $z\rightarrow-m$ with
$m>0$ is as follows.

\begin{figure}[H]
\includegraphics[width=1\textwidth]{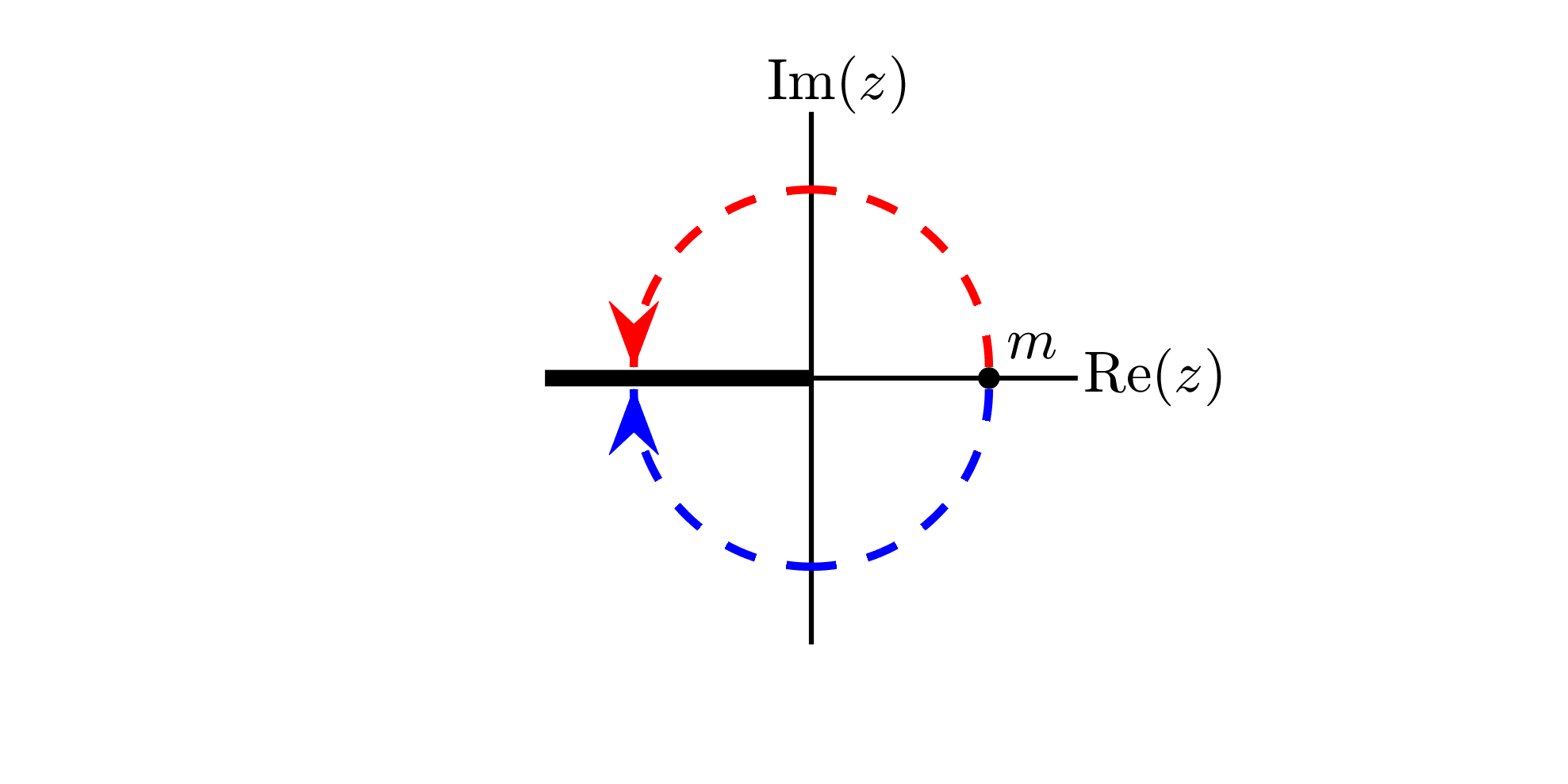} \centering
\caption{\textbf{Arriving at} $\bm{ z=-m}$. In this figure we plot the two
ways we can transform from $z=m$ to $z=-m$ in the complex $z$ plane. There is
a cut along the negative real axis to ensure that $z^{t}$ and $\ln(z)$ are
well-defined functions of $z$.}
\label{fig:arrive}
\end{figure}

With $i=\sqrt{-1}$, we set
\begin{equation}
z=m\exp(i\theta)
\end{equation}
with $\theta$ real and lying in the range $-\pi<\theta<\pi$. We then take
$\theta$ from $0$ to either $-\pi$ or $\pi$, corresponding to $z$ moving on a
semicircular arc around the origin in the complex plane, from $z=m$, and
arriving at either $z=-m-i0$ or $z=-m+i0$, respectively. We illustrate this
procedure in Figure \ref{fig:arrive}.


We find that under this procedure, Eq. (\ref{equiv x'=Lx}) becomes
\begin{equation}
x(t)=m^{t}\exp(-i\pi t)a\quad\text{or}\quad x(t)=m^{t}\exp(i\pi t)a
\label{complex solns}
\end{equation}
while Eq. (\ref{equiv x'=Lx ode}) becomes
\begin{equation}
\frac{dx(t)}{dt}=\left[  \ln(m)-i\pi\right]  x(t)\quad\text{or}\quad
\frac{dx(t)}{dt}=\left[  \ln(m)+i\pi\right]  x(t). \label{complex odes}
\end{equation}
The two solutions in Eq. (\ref{complex solns}) are, for general $t$, both
\textit{complex}, and indeed are \textit{complex conjugates} of each other.
This is also reflected in the ordinary differential equations for $x(t)$ in
Eq. (\ref{complex odes}); the coefficients of $x(t)$, on the right-hand sides
are also complex conjugates of each other.

We can write the two solutions in Eq. (\ref{complex solns}) as ${x(t)=m^{t}
\exp(\mp i\pi t)a=m^{t}\left[  \cos(\pi t)\mp i\sin(\pi t)\right]  a}$ and for
general $t$ these have both real and imaginary parts, i.e., they are
intrinsically complex. However, if we set $t=n$ ($=0,1,2,...$) then the
imaginary parts vanish, and both solutions reduce to $x(n)=(-1)^{n}m^{n}a$,
precisely coinciding with the exact result in Eq. (\ref{subs l=-m}).

From the above example, we cautiously infer the following about exactly
soluble discrete-time problems that exhibit oscillatory behaviour (such as
when $x_{n+1}$ has a different sign to $x_{n}$).

\begin{enumerate}
\item The procedure of obtaining an equivalent continuous-time solution,
$x(t)$, by replacing $n$ by $t$ in the exact discrete-time solution, $x_{n}$,
continues to work, in the sense that $x(n)$ precisely coincides with $x_{n}$.

\item The way the continuous-time \textit{equivalent} solution, $x(t)$,
reproduces the rapid changes in $x_{n}$ is to become \textit{complex}.

\item The differential equation that $x(t)$ obeys generally has complex coefficients.

\item Since there is no reason that either $i$ or $-i$ is `adopted' by the
continuous solution, there are two equivalent continuous-time solutions that
precisely reproduce the value of $x_{n}$ when $t=n$. These solutions are
complex conjugates of each other and have an equal status, with neither
preferable to the other. The two solutions also satisfy differential
equations whose coefficients are complex conjugates of each other.

\item When $t=n$ the imaginary parts of the two equivalent continuous-time
solutions precisely vanish, but for general $t$ the imaginary parts are non-zero.
\end{enumerate}


\subsection{Fibonacci second-order problem}

\label{Fibonacci Section}

We are now in a position to return to the Fibonacci problem that we discussed
at the beginning of this section.

The discrete-time equation is
\begin{equation}
x_{n+2}=x_{n+1}+x_{n}\hspace{0.25cm}\text{with}\hspace{0.25cm}x_{0}
=0\hspace{0.25cm}\text{and}\hspace{0.25cm}x_{1}=1.
\end{equation}
The solution is
\begin{equation}
x_{n}=c\left[  R^{n}-\left(  -\frac{1}{R}\right)  ^{n}\right]  \hspace
{0.25cm}\text{with}\hspace{0.25cm} c=\frac{1}{\sqrt{5}} \hspace{0.25cm}
\text{and} \hspace{0.25cm} R=\frac{1+\sqrt{5}}{2}. \label{Fibonacci soln}
\end{equation}
To obtain the equivalent continuous-time solution, $x(t)$, from the above
expression for $x_{n}$ we follow the approach of Section
\ref{Simple Osc Section}. In particular, we assume the term $\left(
-1/R\right)  ^{n}$, that is present in Eq. (\ref{Fibonacci soln}), had its
origin in the expression $z^{-n}$.

Focussing on $z^{-n}$ as a function of the complex variable $z$, we now
substitute $t$ for $n$ and obtain $z^{-t}$. As in the previous section, we
move $z$, in the complex plane, now from $z=R$ to either $z=-R-i0$ or
$z=-R+i0$, in a similar manner to Figure 6. This leads to $\left(
-1/R\right)  ^{n}$ in Eq. (\ref{Fibonacci soln}) giving rise, in continuous
time, to the two expressions $\exp(\pm i\pi t)R^{-t}$, which are complex
conjugates of each other. It then follows that the continuous-time solutions,
that are equivalent to $x_{n}$, are
\begin{align}
x(t)  &  =c\big(R^{t}-\exp(\pm i\pi t)R^{-t}
\big)\nonumber\label{Fibonacci cont}\\
& \nonumber\\
&  =c\Big(\left[  R^{t}-R^{-t}\cos(\pi t)\right]  \mp iR^{-t}\sin(\pi t)\Big)
\end{align}
with $c$ and $R$ given in Eq. (\ref{Fibonacci soln}). In Figure 7 we have
plotted the solution $x(t)=c\big(R^{t}-\exp(-i\pi t)R^{-t}\big)$ against $t$.


\begin{figure}[H]
\includegraphics[width=1\textwidth]{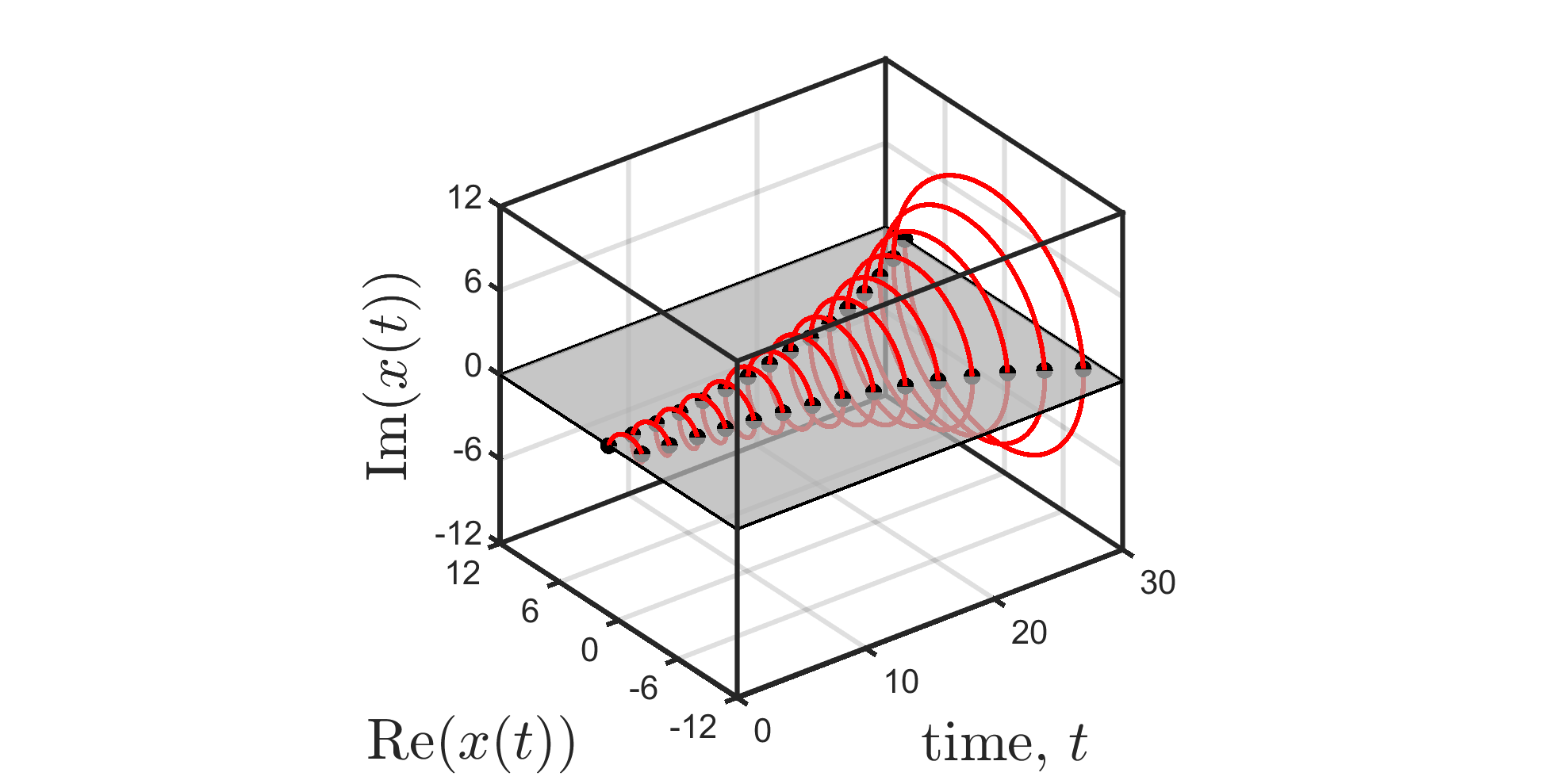} \centering
\caption{\textbf{Complex $\bm{x(t)}$ in the Fibonacci problem}. In this figure
we plot the complex form of $x(t)$ that arises in the case of the dynamics
associated with the Fibonacci numbers, and which is one of the two solutions
in Eq. (\ref{Fibonacci cont}). The red curve depicts $x(t)$ as a function of
the time, $t$, and the black dots are values of $x_{n}$ at $n=0,1,2,...,30$.
The $x_{n}$ are purely real and lie in the shaded plane corresponding to zero
imaginary part.}
\end{figure}

In Appendix E we show that $x(t)$, as given by Eq. (\ref{Fibonacci cont}),
obeys an ordinary differential equation with complex coefficients, and an
initial condition that involves the first derivative also being complex:
\begin{gather}
\frac{d^{2}x(t)}{dt^{2}}\mp i\pi\frac{dx(t)}{dt}-\ln(R)\left[  \ln(R)\mp
i\pi\right]  x(t)=0\\
\nonumber\\
x(0)=0,\quad\quad\left.  \frac{dx(t)}{dt}\right\vert _{t=0}=c\left[
2\ln(R)\mp i\pi\right]  .
\end{gather}

In principle, we could extend the analysis of exactly soluble
oscillatory systems to `close' cases that are not exactly soluble, as we have
done in previous sections. We leave this to future work.


\section{Discussion and Conclusion}

In this work we have established connections between discrete-time dynamical
systems and their continuous-time counterparts.

We proceeded by first looking at both time-homogeneous and time-inhomogeneous
problems that have closed-form solutions of the discrete-time equation,
$x_{n}$, for $n=0,1,2,...$. By the substitution $n\rightarrow t$ we converted
these exact solutions into continuous-time solutions, $x(t)$, that are defined
for $t\geq0$. By construction, $x(t)$ will precisely coincide with $x_{n}$
when $t=n$. What is not obvious is the relation between the difference
equation that $x_{n}$ obeys and the differential equation that $x(t)$ obeys -
the latter found by differentiating $x(t)$ and eliminating initial data. When
we have adopted such a procedure, we have found that there are two major cases:

\noindent$\bullet$ $x_{n}$ changes smoothly, in the sense that no part of
$x_{n}$ rapidly changes sign (i.e., no part of $x_{n}$ has behaviour such as
$(-m)^{n}$ for some $m$)

\noindent$\bullet$ $x_{n}$ has parts that rapidly change sign, e.g., as
$(-m)^{n}$ for some $m$.

\bigskip

\noindent We discuss these two cases separately.

\begin{enumerate}
\item $x_{n}$ changes smoothly

\begin{enumerate}
\item \label{part 1}In the cases considered, we have found that 
time-homogeneous difference equations lead to time-homogeneous differential
equations, and time-inhomogeneous difference equations lead to 
time-inhomogeneous differential equations.

\item Given that there is a reasonably unambiguous identification of
corresponding terms in the difference and differential equations in the cases
considered, we have found that in some discrete-time equations, coefficients
of particular terms are $x$ dependent, but in the differential equation these
coefficients are constants ($x$ independent). This indicates different
functional forms for corresponding coefficients, and for some parameter
regions this means the $x$ dependent coefficient in the difference equation
can vary appreciably with $x$, with no such variation in the differential equation.
\end{enumerate}

\item $x_{n}$ is oscillatory \label{part ii}

In this case, we saw an example where the solution of a discrete-time equation
involved the term $(-m)^{n}$ with $m>0$. Such a term changes sign every
generation. To define a continuous $t$ solution in such a case, we started
with the function $z^{t}$ and had two ways, in the complex $z$ plane, to
unambiguously make $z$ negative. This led to two equal-status solutions that
are complex conjugates of each other. The associated differential equations have
coefficients that are also complex conjugates of each other.
\end{enumerate}

\bigskip

There are a number of comments we can make about the two cases.

\bigskip

For Point 1(a), it was not obvious, at the outset, that a time-homogeneous
differential equation would arise from a time-homogeneous difference equation.
Solutions of a difference equation for $x_{n}$ can depend nontrivially on the
discrete time $n$ (see, e.g., Eq. (\ref{genetics soln})). Promoting $n$, in a
discrete-time solution, to a continuous time variable, $t$, leads to a
possibly nontrivial function $x(t)$. The derivative, $dx(t)/dt$, generally
contains time-dependent terms and generally depends on initial data. We
proceeded by eliminating the initial data from $dx(t)/dt$ and in the cases
considered, this always left a differential equation with time-independent
coefficients. Alternatively, replacing all time-dependent terms in $dx(t)/dt$
by functions of $x(t)$ always led to a differential equation that did not have
coefficients depending on initial data. It is conceivable that either time
dependent coefficients or initial data might arise in the differential
equation. Neither of these possible outcomes occurred. It would be interesting
to have a general understanding of this.

\bigskip

Point 1(a) has the implication that there can be enormous growth in $x_{n}$
with $n$ and yet there still is a continuous-time solution, $x(t)$, that
precisely yields $x(t=n)=x_{n}$. This can be a starting point for an
approximation scheme, for a modification of the equation obeyed by $x_{n}$,
which cannot be solved exactly. We have presented an approach, based on an
approximation of a term in such an equation by an integral, and a continuous
time solution can be found that may capture a lot of the behaviour of $x_{n}$,
and so be `close' to $x_{n}$ for all $n$. This has the potential to allow
accurate solution of problems with extremely rapid growth or decline.

\bigskip

Point 1(b) indicates potentially fundamental differences in the structure of
discrete-time and continuous-time problems. We had, in a discrete-time example, a
coefficient of $x_{n}(1-x_{n})$ of $s/\left(  1+sx_{n}\right)  $ (see Section
\ref{Exact analysis genetics section}). The corresponding coefficient, of
$x(t)[1-x(t)]$ in continuous time, is $\ln\left(  1+s\right)  $. Thus only at
first order in $s$ do $s/\left(  1+sx_{n}\right)  $ and $\ln\left(
1+s\right)  $ agree, but beyond this the $x_{n}$ dependence of the discrete-time 
coefficient indicates a real difference to the continuous-time
coefficient. In situations where a coefficient in the difference equation
fluctuates (see e.g., \cite{jensen1969}, and \cite{Gillespie}) this may lead
to differences when evolution occurs in discrete time as opposed to continuous time.

\bigskip

We have taken the viewpoint in this work that continuous-time dynamics is
derived from discrete time. From a different perspective, we could take the
continuous-time solution, $x(t)$, as describing a system with genuinely
continuous time (in a biological context this would correspond to overlapping
generations). Under such an interpretation, the approach presented in this
work allows us to see the relation of two problems with similar dynamical
variables (e.g., numbers/frequencies), but evolving in discrete or continuous
time. Furthermore, the relation is not restricted to regimes of small parameter values.

\bigskip

In Point 2, the `complexification' of the solutions changes the picture of
$x(t)$, as an interpolation of $x_{n}$ to non-discrete $n$. Now, $x_{n}$ with
an oscillatory part does not yield a continuous-time solution that directly
interpolates the $x_{n}$. Rather, in such a case, when $t$ is not an integer,
$x(t)$ is \textit{complex}, as is illustrated in Figure $7$.

In the case of the Fibonacci numbers, as considered in Section
\ref{Fibonacci Section}, as the discrete time $n$ increases, the `problematic'
$(-R)^{-n}$ term makes an increasingly small contribution to the full form of
$x_{n}$. In discrete time the term $(-R)^{-n}$ seems simply to correct the
exponentially growing term ($R^{n}$), to make the final result integer-valued.
Thus a term of numerically small magnitude can have a very large effect on the
nature of the equivalent continuous-time solution.

\bigskip

The above are basic considerations of the results we have presented. 
They naturally lead on to several questions. The most direct is whether there is a
procedure that goes \textit{directly} from a difference equation to an
equivalent differential equation.

Beyond this, all the problems we have considered are deterministic, so
our results apply to deterministic dynamical systems. We recognise that
many biological systems exhibit stochasticity. An interesting way
forward is thus to see whether the approach presented can be extended to
stochastic dynamical systems, where random terms corresponding to noise
enter the dynamical equations. For the deterministic systems considered,
we used the notion of \textit{equivalent solutions}, where the discrete-time
and continuous-time problems had solutions that agreed precisely at the
discrete times. With difference equations containing noise, equivalence
to some form of stochastic differential equation or other representation
would perhaps need to be softened to a \textit{statistical equivalence}. We note
that in the context of Fibonacci numbers, the generalisation to include
random coefficients has been made by \cite{Embree} and \cite{Viswanath}, and this
work may be a relevant starting point for future investigations.

\newpage

\appendix

\begin{center}
{\large \textbf{APPENDICES}}
\end{center}

\section{Numerically illustrating the relation between an exactly soluble
discrete-time problem and its continuous-time analogue}

In this appendix we numerically illustrate the relation between an exactly
soluble discrete-time problem and its continuous-time analogue, for the
particular case considered in Section \ref{Exact pop, inhomog}, where 
time-inhomogeneous population size change was investigated.

We use the values $t=n=20$, along with the choices of $a$ and $b$ used in
Figure \ref{fig:simple inhomog}.

\begin{figure}[H]
\includegraphics[width=1\textwidth]{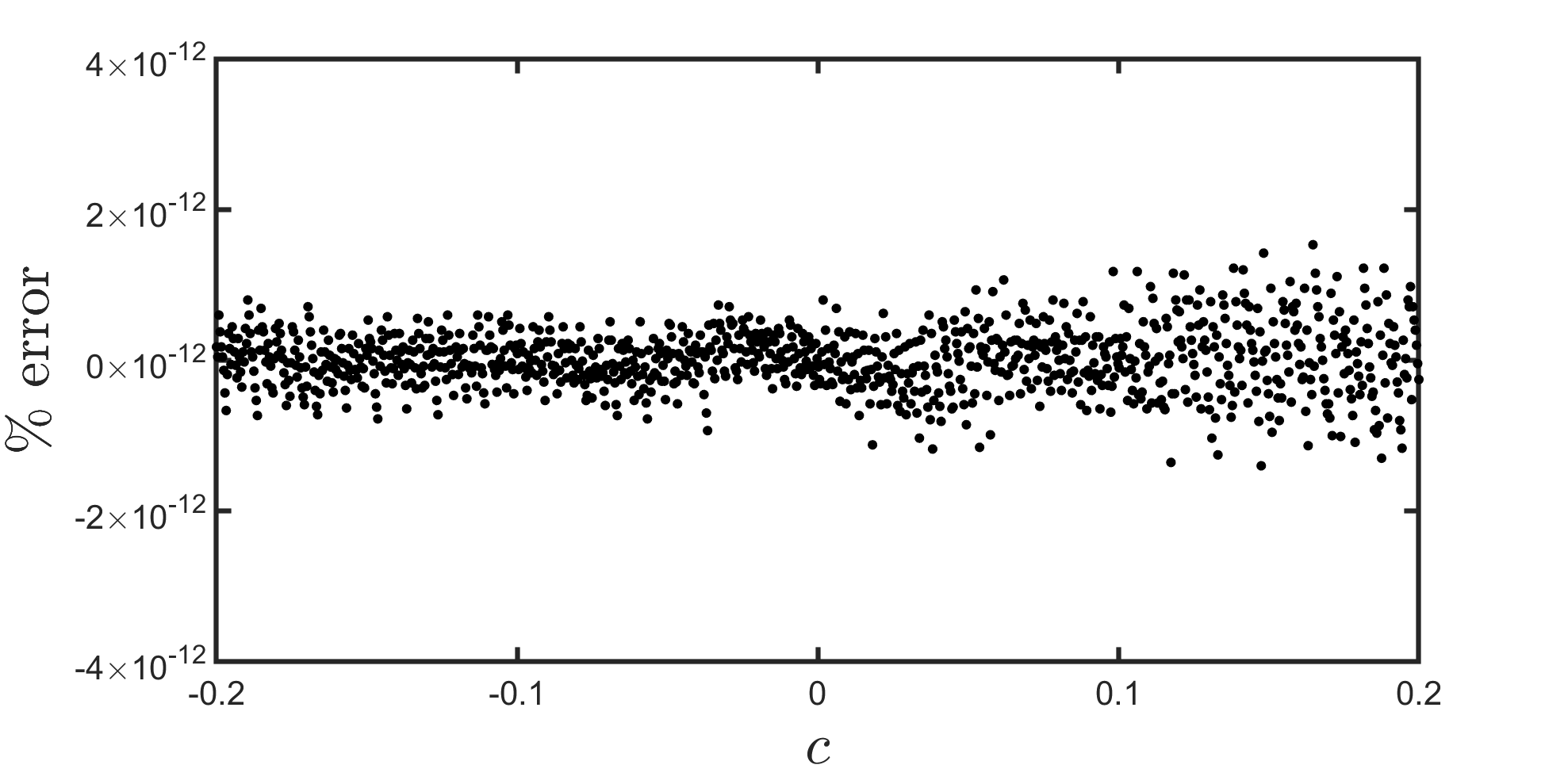} \centering
\caption{\textbf{Numerically calculated error of the continuous-time and 
discrete-time solutions for an exactly soluble form of $\bm{r(k)}$.} In this figure,
for the special, exactly soluble, form of $r(k)$ given by $r(k)=\exp
{(b+ck)}-1$ (Eq. (\ref{r special})), we plot the numerically calculated error
between the exact continuous solution, $x(t)=\exp\left(  bt+c\frac{t(t+1)}
{2}\right)  a$, (Eq. (\ref{special cont t soln})) and the result for $x_{n}$
obtained by iteration of Eq. (\ref{simple inhomog}). Analytically, we expect
that for this form of $r(k)$ the discrete-time and continuous-time solutions
precisely coincide at $t=n$. The figure was obtained by numerically computing
the percentage error for $t=n=20$, namely $[x(20)-x_{20}]/x_{20}\times100$,
for a range of $c$ values that included the value used for Figure \ref{fig:simple inhomog}. The
values of the parameters $a$ and $b$ used were identical to those used in
Figure \ref{fig:simple inhomog}.}
\label{fig:error appendix}
\end{figure}

We numerically compared $x(20)$, as calculated from the analytical formula in
Eq. (\ref{special cont t soln}), and the value of $x_{20}$ that was obtained
from direct iteration of Eq. (\ref{simple inhomog}) (we used iteration
because, in general, iteration is the only way to find the $x_{n}$). The
results are plotted in Figure \ref{fig:error appendix}.


The numerically calculated error between the exact continuous-time solution
and the discrete-time solution that was calculated by iteration is, as shown
in Figure \ref{fig:error appendix}, of the order of $1$ part in $10^{14}$.
This gives an indication of the sort of errors that arise from numerical
evaluation.


\section{Representing the dynamics in the genetics model}

In this appendix we determine the dynamics of a population of haploid asexual
organisms that evolve in discrete generations, as considered in Section
\ref{Genetics section} of the main text.

Individuals carry one of two genes, labelled $A$ and $B$, and in the adults of
generation $n$, with $n=0,1,2,...$, the number of carriers of the $A$ gene
is $N_{n}^{(A)}$ and that of the $B$ gene is $N_{n}^{(B)}$.

The life cycle is: (i) adults produce offspring, and if they contain a
mutation then they have a gene different from the parental gene; (ii) the
adults die after reproduction; (iii) the offspring undergo number regulation,
leaving $N$ individuals who constitute the adults of the next generation. We
assume $N$ is sufficiently large that the dynamics can be treated as deterministic.

The carrier of an $A$ gene produces a total of $\alpha$ offspring, while a
carrier of a $B$ gene produces a total of $\beta$ offspring.

Let $u$ denote the probability that an $A$ gene in a parent mutates to a $B$
gene in an offspring, while $v$ denotes the corresponding probability that a
$B$ gene mutates to an $A$ gene.

After reproduction, we write the number of offspring carrying the $A$ and $B$
genes as $N_{n}^{(A)\ast}$ and $N_{n}^{(B)\ast}$, respectively. We have
\begin{align}
N_{n}^{(A)\ast}  &  =(1-u)\alpha N_{n}^{(A)}+v\beta N_{n}^{(B)}\nonumber\\
& \\
N_{n}^{(B)\ast}  &  =(1-v)\beta N_{n}^{(B)}+u\alpha N_{n}^{(A)}.\nonumber
\end{align}
After number regulation, the numbers of adults carrying the different genes in
the next generation are given by
\begin{align}
N_{n+1}^{(A)}  &  =R_{n}N_{n}^{(A)\ast}=R_{n}\left[  (1-u)\alpha N_{n}
^{(A)}+v\beta N_{n}^{(B)}\right] \nonumber\\
& \label{Nn eqs}\\
N_{n+1}^{(B)}  &  =R_{n}N_{n}^{(B)\ast}=R_{n}\left[  (1-v)\beta N_{n}
^{(B)}+u\alpha N_{n}^{(A)}\right] \nonumber
\end{align}
where $R_{n}$ is a `number regulation factor' (a function of $N_{n}^{(A)}$ and
$N_{n}^{(B)}$ that ensures the sum of the numbers of $A$ and $B$ gene carriers
equals $N$).

Let
\begin{equation}
x_{n}=\frac{N_{n}^{(A)}}{N_{n}^{(A)}+N_{n}^{(B)}}
\end{equation}
denote the frequency of adult carriers of the $A$ gene in generation $n$, with
the corresponding frequency of $B$ gene carriers given by $1-x_{n}$. Then from
Eq. (\ref{Nn eqs}) we obtain
\begin{equation}
x_{n+1}=\frac{(1-u)\alpha N_{n}^{(A)}+v\beta N_{n}^{(B)}}{(1-u)\alpha
N_{n}^{(A)}+v\beta N_{n}^{(B)}+(1-v)\beta N_{n}^{(B)}+u\alpha N_{n}^{(A)}}
\end{equation}
which can be written as
\begin{equation}
x_{n+1}=\frac{(1-u)\alpha x_{n}+v\beta(1-x_{n})}{\alpha x_{n}+\beta(1-x_{n}
)}. \label{xn+1=... appendix}
\end{equation}
We can express this equation in the equivalent form
\begin{equation}
x_{n+1}=x_{n}+\sigma(x_{n})x_{n}(1-x_{n})-U(x_{n})x_{n}+V(x_{n})(1-x_{n})
\end{equation}
where
\begin{align}
\sigma(x)  &  =\frac{\alpha-\beta}{\beta+(\alpha-\beta)x},\nonumber\\
& \nonumber\\
U(x)  &  =\frac{u\alpha}{\beta+(\alpha-\beta)x},\label{3}\\
& \nonumber\\
V(x)  &  =\frac{v\beta}{\beta+(\alpha-\beta)x}.\nonumber
\end{align}
The three quantities in Eq. (\ref{3}) can be expressed in terms of the
selection coefficient $s$ defined by
\begin{equation}
\frac{\alpha}{\beta}=1+s.
\end{equation}
We then find
\begin{align}
\sigma(x)  &  =\frac{1}{1+sx},\nonumber\\
& \nonumber\\
U(x)  &  =\frac{u(1+s)}{1+sx},\\
& \nonumber\\
V(x)  &  =\frac{v}{1+sx}.\nonumber
\end{align}


\section{Solution of the dynamics in the genetics model}

In this appendix we derive the exact solution of the genetics problem, as
described in Section \ref{Exact analysis genetics section} of the main text
and Appendix B.

To begin, we write $x\equiv x_{n}$ and $x^{\prime}\equiv x_{n+1}$ and Eq.
(\ref{xn+1=... appendix}), which is equivalent to Eq. (\ref{x n+1 genetics})
of the main text, takes the form
\begin{equation}
x^{\prime}=\frac{(1-u)\alpha x+v\beta(1-x)}{\alpha x+\beta(1-x)}.
\end{equation}
Let $\hat{x}$ be the equilibrium solution that solves
\begin{equation}
\hat{x}=\frac{(1-u)\alpha\hat{x}+v\beta(1-\hat{x})}{\alpha\hat{x}+\beta
(1-\hat{x})}.
\end{equation}
We find the solution that lies in the range $[0,1]$ is given by
\begin{align}
\hat{x} =  &  \frac{(1-u)\alpha-(1+v)\beta}{2(\alpha-\beta)}\nonumber\\
& \nonumber\\
&  +\frac{\sqrt{\left[  (1-u)\alpha-(1-v)\beta\right]  ^{2}+4uv\alpha\beta}
}{2(\alpha-\beta)}.
\end{align}
Then we can write
\begin{equation}
x^{\prime}-\hat{x}=\frac{(1-u)\alpha x+v\beta(1-x)}{\alpha x+\beta(1-x)}
-\frac{(1-u)\alpha\hat{x}+v\beta(1-\hat{x})}{\alpha\hat{x}+\beta(1-\hat{x})}.
\end{equation}
We set
\begin{equation}
y=\frac{1}{x-\hat{x}}
\end{equation}
and then find
\begin{align}
y^{\prime}  &  =\left(  \tfrac{(1-u)\alpha x+v\beta(1-x)}{\alpha x+\beta
(1-x)}-\tfrac{(1-u)\alpha\hat{x}+v\beta(1-\hat{x})}{\alpha\hat{x}+\beta
(1-\hat{x})}\right)  ^{-1}\nonumber\\
& \nonumber\\
&  =\tfrac{(\alpha\hat{x}+\beta-\beta\hat{x})^{2}}{\alpha\beta(1-u-v)}
y+\tfrac{(\alpha-\beta)(\alpha\hat{x}+\beta-\beta\hat{x})}{\alpha\beta
(1-u-v)}\nonumber\\
& \nonumber\\
&  \equiv\lambda y+\mu
\end{align}
where
\begin{equation}
\lambda=\frac{(\alpha\hat{x}+\beta-\beta\hat{x})^{2}}{\alpha\beta
(1-u-v)}\hspace{0.25cm}\text{and}\hspace{0.25cm}\mu=\frac{(\alpha
-\beta)(\alpha\hat{x}+\beta-\beta\hat{x})}{\alpha\beta(1-u-v)}.\nonumber
\end{equation}
The solution is
\begin{align}
y_{n}  &  =\left\{
\begin{array}
[c]{ll}
y_{0}, & n=0\\
& \\
\lambda^{n}y_{0}+\sum_{j=0}^{n-1}\lambda^{j}\mu, & n=1,2,...
\end{array}
\right. \nonumber\\
& \nonumber\\
&  =\left\{
\begin{array}
[c]{ll}
y_{0}, & n=0\\
& \\
\lambda^{n}y_{0}+\frac{1-\lambda^{n}}{1-\lambda}\mu, & n=1,2,...
\end{array}
\right.
\end{align}
leading to the exact solution for $x_{n}$:
\begin{equation}
x_{n}=\hat{x}+\frac{1}{\lambda^{n}y_{0}+\frac{1-\lambda^{n}}{1-\lambda}\mu
}=\hat{x}+\frac{1}{\frac{\lambda^{n}}{x_{0}-\hat{x}}+\frac{1-\lambda^{n}
}{1-\lambda}\mu}.
\end{equation}
In terms of the selection coefficient $s$ defined by
\begin{equation}
\alpha/\beta=1+s
\end{equation}
we have
\begin{align}
\hat{x} =  &  \frac{(1-u)(1+s)-(1+v)}{2s}\nonumber\\
& \nonumber\\
&  +\frac{\sqrt{\left[  (1-u)(1+s)-(1-v)\right]  ^{2}+4uv(1+s)}}{2s}
\end{align}
along with
\begin{equation}
\lambda=\frac{(1+s\hat{x})^{2}}{(1+s)(1-u-v)}\hspace{0.25cm}\text{and}
\hspace{0.25cm}\mu=\frac{s(1+s\hat{x})}{(1+s)(1-u-v)}.\nonumber
\end{equation}


\section{Solution of the time-inhomogeneous dynamics in the purely selective
genetics model}

In this appendix we determine the solution of the difference equation that
occurs in Eq. (\ref{xn+1 genetics inhomog}) of the main text.

We find it more convenient to carry out some of the algebraic manipulations in
terms of the (discrete) time-dependent $\alpha$ and $\beta$, prior to
expressing their ratio in terms of a time-dependent selection coefficient.

In this case we take the time-dependent mutation-free analogue of Eq.
(\ref{x n+1 genetics}) to be
\begin{equation}
x_{n+1}=\frac{\alpha_{n+1}x_{n}}{\alpha_{n+1}x_{n}+\beta_{n+1}(1-x_{n} )}.
\label{D1}
\end{equation}
We write this as
\begin{align}
\frac{1}{x_{n+1}}  &  =\frac{\alpha_{n+1}x_{n}+\beta_{n+1}(1-x_{n})}
{\alpha_{n+1}x_{n}}\nonumber\\
& \nonumber\\
&  =1-\frac{\beta_{n+1}}{\alpha_{n+1}}+\frac{\beta_{n+1}}{\alpha_{n+1}}
\frac{1}{x_{n}}
\end{align}
or
\begin{equation}
\left(  \frac{1}{x_{n+1}}-1\right)  =\frac{\beta_{n+1}}{\alpha_{n+1}}\left(
\frac{1}{x_{n}}-1\right)  .
\end{equation}
The solution for $n>0$ is
\begin{align}
\left(  \frac{1}{x_{n}}-1\right)  =  &  \frac{\beta_{n}}{\alpha_{n}}
\frac{\beta_{n-1}}{\alpha_{n-1}}\cdots\frac{\beta_{1}}{\alpha_{1}}\left(  \frac{1}
{x_{0}}-1\right) \nonumber\\
& \nonumber\\
=  &  \exp\left(  -\sum_{k=0}^{n-1}\ln\left(  \frac{\alpha_{k+1}}{\beta_{k+1}
}\right)  \right)  \times\left(  \frac{1}{x_{0}}-1\right)
\end{align}
thus
\begin{equation}
x_{n}=\frac{1}{\exp\left(  -\sum_{k=0}^{n-1}\ln\left(  \frac{\alpha_{k+1}
}{\beta_{k+1}}\right)  \right)  \left(  \frac{1}{x_{0}}-1\right)  +1}.
\end{equation}
With
\begin{equation}
\frac{\alpha_{n}}{\beta_{n}}=1+s(n)
\end{equation}
we have
\begin{equation}
x_{n}=\left\{
\begin{array}
[c]{ll}
x_{0}, & n=0\\
& \\
\tfrac{x_{0}}{\left(  1-x_{0}\right)  \exp\left(  -\sum_{k=0}^{n-1}\ln\left(
1+s(k+1)\right)  \right)  +x_{0}}, & n\geq1.
\end{array}
\right.
\end{equation}
Similarly to the population growth problem, we shall approximate $\ln\left(
1+s\left(  k+1\right)  \right)  $ by an \textit{integral} (again using the
approximate mid-point integration rule in the opposite direction to usual):
\begin{equation}
\ln\left(  1+s\left(  k+1\right)  \right)  \simeq\int_{k}^{k+1}\ln\left(
1+s\left(  z+\frac{1}{2}\right)  \right)  dz \label{approx appendix}
\end{equation}
then
\begin{equation}
x_{n}\simeq\left\{
\begin{array}
[c]{ll}
x_{0}, & n=0\\
& \\
\frac{x_{0}}{\left(  1-x_{0}\right)  \exp\left(  -I\right)  +x_{0}}, & n\geq1
\end{array}
\right.
\end{equation}
where
\begin{equation}
I=\sum_{k=0}^{n-1}\int_{k}^{k+1}\ln\left(  1+s\left(  z+\tfrac{1}{2}\right)
\right)  dz
\end{equation}
and we can write this for $n=0,1,2,...$ as
\begin{equation}
x_{n}\simeq\frac{x_{0}}{\left(  1-x_{0}\right)  \exp\left(  -\int_{0}^{n}
\ln\left(  1+s\left(  z+\frac{1}{2}\right)  \right)  dz\right)  +x_{0}}.
\end{equation}
This leads us to define
\begin{equation}
x^{(\mathrm{app})}(t)=\frac{x_{0}}{\left(  1-x_{0}\right)  \exp\left(
-\int_{0}^{t}\ln\left(  1+s\left(  z+\frac{1}{2}\right)  \right)  dz\right)
+x_{0}}
\end{equation}
which satisfies
\begin{equation}
\frac{dx^{(\mathrm{app})}(t)}{dt}=\ln\left(  1+s\left(  t+\tfrac{1}{2}\right)
\right)  \times x^{(\mathrm{app})}(t)\left[  1-x^{(\mathrm{app})}(t)\right]  .
\label{ode special appendix}
\end{equation}
The original difference equation, Eq. (\ref{D1}), can be written as
\begin{equation}
x_{n+1}-x_{n}=\frac{s(n+1)}{1+s(n+1)x_{n}}x_{n}(1-x_{n}). \label{equiv D1}
\end{equation}
Comparing the previous two equations, we arrive at the (generally approximate)
mapping
\begin{equation}
\frac{s(n+1)}{1+s(n+1)x_{n}}\rightarrow\ln\left(  1+s\left(  t+\tfrac{1}
{2}\right)  \right)
\end{equation}
and as in the time-homogeneous case, a frequency-dependent coefficient in the
discrete-time problem becomes a frequency independent coefficient in
continuous time. Furthermore, the approximation in Eq. (\ref{approx appendix})
is \textit{exact} when $\ln\left(  1+s\left(  t+\frac{1}{2}\right)  \right)  $
is a linear function of $t$. In this case, Eq. (\ref{ode special appendix})
will have a solution that precisely agrees with the solution of Eq.
(\ref{equiv D1}) when $t=n$, irrespective of the size of $s(t)$.


\section{Derivation of the differential equation in a second-order problem}

In this appendix we derive the differential equation obeyed by the function of
$t$ given in Eq. (\ref{Fibonacci cont}) of the main text. The function is
\begin{equation}
x(t)=c\left(  R^{t}-e^{\pm i\pi t}R^{-t}\right)  \label{full F soln appendix}
\end{equation}
where $c=1/\sqrt{5}$ and $R=(1+\sqrt{5})/2$.

The function in Eq. (\ref{full F soln appendix}) arose from a second-order
difference equation, and to facilitate the derivation of the differential
equation it obeys, we write
\begin{equation}
x(t)=AR^{t}+BS^{-t} \label{F soln appendix}
\end{equation}
where we treat $A$ and $B$ as arbitrary constants that are determined by
initial data, and shall look for a differential equation that is independent
of $A$ and $B$.

We have
\begin{align}
\frac{dx}{dt}  &  =A\ln(R)R^{t}-B\ln(S)S^{-t}\\
& \nonumber\\
\frac{d^{2}x}{dt^{2}}  &  =A\ln^{2}(R)R^{t}+B\ln^{2}(S)S^{-t}.
\end{align}
From these equations we obtain
\begin{align}
A  &  =\frac{1}{R^{t}}\frac{1}{\ln(R)+\ln(S)}\left(  \frac{d^{2}x}{dt^{2}}
+\ln(S)\frac{dx}{dt}\right) \\
& \nonumber\\
B  &  =\frac{1}{S^{t}}\frac{1}{\ln(R)+\ln(S)}\left(  \frac{d^{2}x}{dt^{2}}
-\ln(R)\frac{dx}{dt}\right)  .
\end{align}
Substituting these into Eq. (\ref{F soln appendix}) yields the time-homogeneous equation
\begin{equation}
\frac{d^{2}x}{dt^{2}}-\left[  \ln(R)-\ln(S)\right]  \frac{dx}{dt}-\ln
(R)\ln(S)x=0.\nonumber
\end{equation}
Setting $\ln(S)=\ln(R)\mp i\pi$ yields the differential equation that the
function in Eq. (\ref{full F soln appendix}) obeys:
\begin{equation}
\frac{d^{2}x}{dt^{2}}\mp i\pi\frac{dx}{dt}-\ln(R)\left[  \ln(R)\mp
i\pi\right]  x=0 \label{Fib ode appendix}
\end{equation}
From Eq. (\ref{full F soln appendix}) we have
\begin{equation}
x(0)=0\hspace{0.25cm}\hspace{0.25cm}\text{and}\hspace{0.25cm}\hspace
{0.25cm}\left.  \frac{dx(t)}{dt}\right\vert _{t=0}=c\left[  2\ln(R)\mp
i\pi\right]  .
\end{equation}

\newpage

\bibliography{Bibliography}

\end{document}